\documentclass[Afour,sagev,times]{sagej}

\usepackage{moreverb,url}
\usepackage{float}
\usepackage{amsmath}
\usepackage{amssymb}
\usepackage{booktabs}
\usepackage{subcaption}
\usepackage{float}
\usepackage{bm}

\usepackage{hyperref}

\hypersetup{colorlinks,bookmarksopen,bookmarksnumbered,citecolor=red,urlcolor=red}
\DeclareMathOperator*{\plim}{plim\,}
\DeclareMathOperator*{\miota}{M_{\boldsymbol{1}}}

\DeclareMathOperator*{\one}{\boldsymbol{1}}
\DeclareMathOperator*{\bav}{\hat{\beta}_a^{|bav}}
\DeclareMathOperator*{\naive}{\hat{\beta}_a^{naive}}
\DeclareMathOperator*{\siga}{\sigma_a^2}
\DeclareMathOperator*{\sigu}{\sigma_u^2}
\DeclareMathOperator*{\sigbav}{\sigma_{bav}^2}
\DeclareMathOperator*{\gu}{\gamma_u}
\DeclareMathOperator*{\gbav}{\gamma_{bav}}
\DeclareMathOperator*{\bbav}{\beta_{bav}}
\DeclareMathOperator*{\ba}{\beta_{a}}
\DeclareMathOperator*{\bu}{\beta_{u}}
\DeclareMathOperator*{\eone}{\boldsymbol{\epsilon_{1}}}
\DeclareMathOperator*{\etwo}{\boldsymbol{\epsilon_{2}}}
\DeclareMathOperator*{\ay}{\alpha_{y}}

\DeclareMathOperator*{\ama}{\boldsymbol{A^TM_zA}}
\DeclareMathOperator*{\amy}{\boldsymbol{A^TM_zY}}
\DeclareMathOperator*{\amu}{\boldsymbol{A^TM_zU}}

\newcommand\BibTeX{{\rmfamily B\kern-.05em \textsc{i\kern-.025em b}\kern-.08em
T\kern-.1667em\lower.7ex\hbox{E}\kern-.125emX}}

\def\volumeyear{2016}
\begin{document}

\runninghead{Stokes et al}

\title{Causal Simulation Experiments: Lessons from Bias Amplification}

\author{Tyrel Stokes\affilnum{1}, Russell Steele\affilnum{1}, and Ian Shrier\affilnum{2}\affilnum{3}}

\affiliation{\affilnum{1}McGill University, Department of Mathematics and Statistics\\
\affilnum{2}McGill University, Department of Family Medicine\\
\affilnum{3}Centre for Clinical Epidemiology, Lady Davis Institute}

\corrauth{Russell Steele, McGill University
Department of Mathematics and Stistics,
Burnside Hall, Room 1005
805 Sherbrooke Street West
Montreal, Quebec
Canada
H3A 0B9}

\email{russell.steele@mcgill.ca}

\begin{abstract}
Recent theoretical work in causal inference has explored an important class of variables which, when conditioned on, may further amplify existing unmeasured confounding bias (bias amplification). Despite this theoretical work, existing simulations of bias amplification in clinical settings have suggested bias amplification may not be as important in many practical cases as suggested in the theoretical literature. We resolve this tension by using tools from the semi-parametric regression literature leading to a general characterization in terms of the geometry of OLS estimators which allows  us to extend current results to a larger class of DAGs, functional forms, and distributional assumptions. We further use these results to understand the limitations of current simulation approaches and to propose a new framework for performing causal simulation experiments to compare estimators. We then evaluate the challenges and benefits of extending this simulation approach to the context of a real clinical data set with a binary treatment, laying the groundwork for a principled approach to sensitivity analysis for bias amplification in the presence of unmeasured confounding.
\end{abstract}


\maketitle

\section{Introduction}

\label{sec:intro}
\setcounter{page}{1}

\begin{figure}[H]
\begin{center}
\includegraphics[scale=.32]{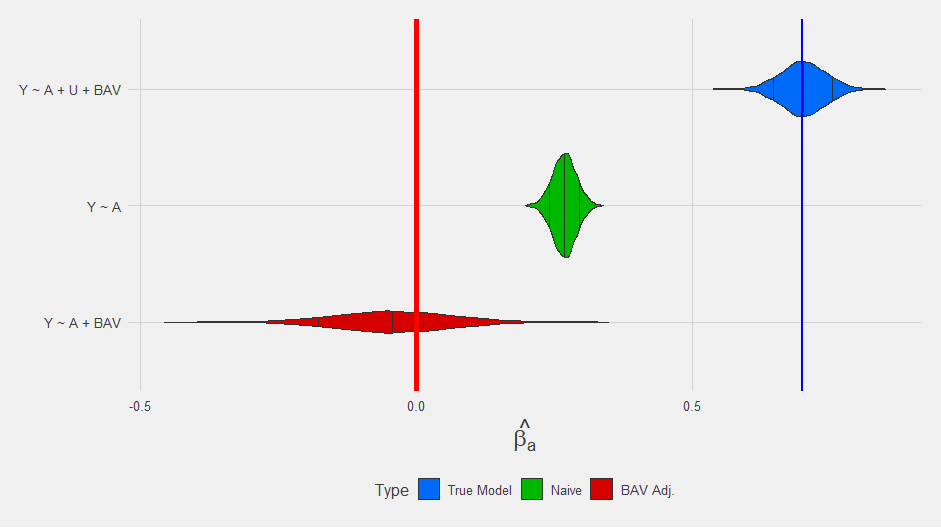} 
\caption{\small Violin plot from simulations from equations \eqref{eq1: big meyers},\eqref{eq2: big meyers}, and \eqref{eq3: big meyers}. There were 5000 replication with $n=5000$. The true effect of interest was $\beta_a =0.7$, represented by the blue line. The confounding effects were $\beta_u=-0.5$ and $\gamma_u=0.59$. The vector of coefficients for $\boldsymbol{BAV}$ on U was $\psi_{bav} = \{-0.55, -0.45, -0.3,  0.30, .25, 0.20, -0.20, 0.20,  -0.15, 0.10 \}$, which in general were larger than the impact of $BAV$ on $A$, $\gamma_{bav} = \{-.1,-.15,-.1,.21,-.2,.3,-.2,-.15,-.2,.075\}$. The red horizontal represents $\hat{\beta_a} = 0$.}\normalsize
\label{fig:machinesim}
\end{center}
\end{figure}

Causal identification strategies aim to condition on a sufficient set of observables such that the potential outcomes are conditionally independent of the treatment of interest  \citep{rubin1974estimating,  rosenbaum1983central,wooldridge2010econometric}. Causal variable selection procedures often assume that at least one subset of the observed variables forms such a sufficient set \citep{witte2018covariate}. The object in causal variable selection then becomes how to separate variables which are necessary for identification of the causal effect from those variables which are extraneous \citep{witte2018covariate,hernan2002} in the interest of reducing estimate variance or covariate dimensionality \citep{greenland2015statistical,witte2018covariate}.\\
\indent In non-experimental observational studies,  we do not have full access to a sufficient set in many realistic settings, and important confounding pathways remain unblocked \citep{VanderWeelePeng_Sensitivity2017,hill_sensitivity}. This is referred to as unmeasured confounding or endogeneity in the statistics and econometrics literatures respectively. However, applied researchers currently rely on variable selection techniques  such as lasso, step-wise, change-in-estimator selection, and outcome and/or treatment oriented approaches \citep{Talbot2019} despite violating their underlying assumptions.\\
\indent The use of variable selection techniques is to avoid conditioning on negligible confounding pathways without introducing meaningful bias to the estimator. In this paper we explore how this intuition can break down under even mild violations of the underlying assumptions, particularly under the threat of bias amplification.\\
First consider data generated from the following directed acyclic graph (DAG) (Figure \ref{fig:DAG_exp}) and set of structural equations:

\begin{figure}[H]
	\centering
	\includegraphics[scale=.33]{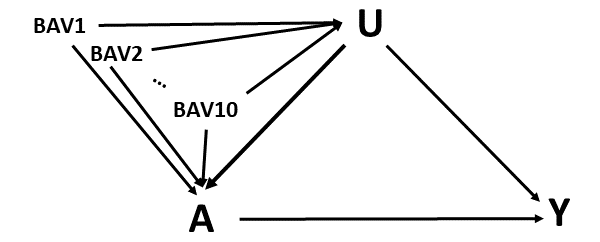}
	\caption{Directed acycic graph (DAG): Meyers (2011) extended to 10 possible bias amplifying variables BAVs.}
	\label{fig:DAG_exp}
\end{figure}
\begin{align}
    \boldsymbol{Y} &= \alpha_y + \boldsymbol{A}\beta_a + \boldsymbol{U}\beta_u + \epsilon_1\label{eq1: big meyers}, \\
    \boldsymbol{A} &= \alpha_a + \boldsymbol{U}\gamma_u + \sum_{i=1}^{10}\boldsymbol{BAV_i}\gamma_{bav_i} + \epsilon_2\label{eq2: big meyers}, \\
    \boldsymbol{U} &= \alpha_u + \sum_{i=1}^{10}\boldsymbol{BAV_i}\psi_{bav_i} + \epsilon_3, \label{eq3: big meyers}
\end{align}
 
 \noindent where $\boldsymbol{Y}$ is the outcome, $\boldsymbol{A}$ is the treatment of interest, $\boldsymbol{U}$ is an unmeasured variable and $\boldsymbol{BAV}$ refers to 10 different potential bias amplifying  variables that are measured and affect both $\boldsymbol{A}$ and $\boldsymbol{U}$ but have no direct effect on $\boldsymbol{Y}$. This model contains one confounding path that cannot be blocked  ($\boldsymbol{A}\leftarrow \boldsymbol{U} \rightarrow \boldsymbol{Y}$) and 10 confounding paths ($\boldsymbol{A}\leftarrow \boldsymbol{BAV_1} \rightarrow \boldsymbol{U} \rightarrow \boldsymbol{Y}$; \dots ; $\boldsymbol{A}\leftarrow \boldsymbol{BAV_{10}} \rightarrow \boldsymbol{U} \rightarrow \boldsymbol{Y}$) that can be blocked by including the measured $\boldsymbol{BAV_i}$ variables in the model. However, including any of these $\boldsymbol{BAV_i}$ might also increase bias (potential bias amplifying variables). Our goal is to find the least biased estimator of the average causal effect of treatment ($\beta_a$).\\
\indent By including more $\boldsymbol{BAV}s$, intuition suggests the remaining unmeasured confounding bias should decrease. However as demonstrated in the bias amplification literature \cite{pearl2012class,pearl2011invited,middleton2016bias} conditioning on confounders may still increase bias. For example, suppose further the 10 observable variables account for 90\% of the variance in the variable $\boldsymbol{U}$ responsible for unmeasured confounding. The blue violin plot in Figure \ref{fig:machinesim} represents the density of the estimates from the true outcome model with the treatment and both measured/unmeasured confounding variables included as regressors. As expected the estimates are approximately normally distributed around the true value $\beta_a =0.7$. The green violin plot represents the biased estimates from the naive model, the simple regression of the outcome, $\boldsymbol{Y}$, on the treatment $\boldsymbol{A}$, which does not include any of the confounders (measured or unmeasured). The red violin plot represents the linear model adjusted for all 10 measured confounders which account for $90\%$ of the unmeasured confounding. The adjusted model performs much worse than the naive model both in terms of bias (0.73 compared to 0.43, interpretable as standard deviations) and variance (standard deviation of 0.1 compared to 0.02). In fact, in 4990 of 5000 simulations the adjusted estimate was farther from the truth than the naive estimate and nearly 65\% of the adjusted estimates had the incorrect effect sign.

The purpose of this paper is to explain why model selection intuition fails us in this case and how we can use a combination of data and simulation approaches to improve model selection. We build upon an emerging theoretical literature exploring a class of variables which can amplify existing unmeasured confounding bias \citep{WOOLDRIDGE2016232,pearl2012class,ding2017instrumental}. This class of variables is potentially very large and common in practical applications. Finally, we discuss possible model selection strategies to minimise bias and variance when unmeasured confounding is believed to be present.\\
\indent We adopt a matrix notation framework to characterize this problem because 1) we can easily generalize to a much larger class of directed acyclic graphs and structural equations than previously studied, 2) it offers a unifying geometric explanation in the context of least squares estimation and 3) it offers a solid foundation for how to build data informed model selection procedures. Finally, we develop a procedure for simulating from a more complete parameter space in a way that respects the underlying amplification process. In addition to lending itself better to articulating and answering causal simulation questions this procedure helps explain why some previous studies have incorrectly concluded that applied investigators need not worry about amplification in practice\cite{meyers2011}. We evaluate the challenges of implementing this approach with a real clinical example with binary treatment.

\section{Problem Formulation}

\begin{figure}[H]
	\centering
\includegraphics[scale=.3]{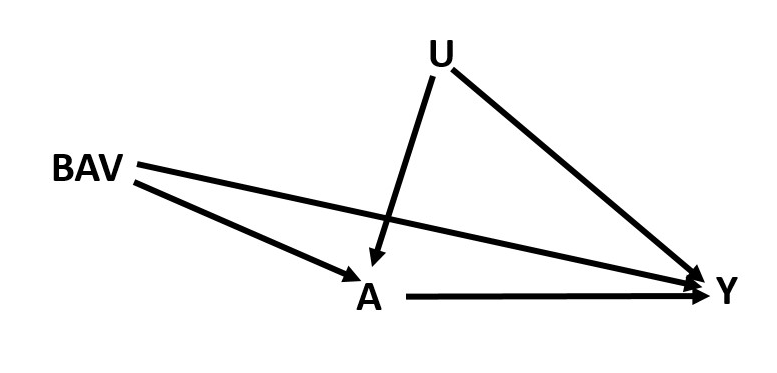}
\caption{DAG: Two Confounding Paths, where $A$ is the treatment of interest, $Y$ is the outcome, $U$ is an unmeasured variable and $BAV$ is a measured variable}
\label{fig:DAG1}
\end{figure}

Figure \ref{fig:DAG1} shows a directed acyclic graph (DAG) for a simpler model containing both measured and unmeasured confounding. Let $\boldsymbol{Y}$ represent the outcome and $\boldsymbol{A}$ is the treatment or variable of interest. Let $\boldsymbol{U}$ be an unmeasured confounding variable that we cannot include in a regression model, but which has a functional relationship with both $\boldsymbol{Y}$ and $\boldsymbol{A}$. The bias amplifying variable ($\boldsymbol{BAV}$) in this DAG is analagous to $\boldsymbol{U}$ in that it is a cause of $\boldsymbol{Y}$ and a cause of $\boldsymbol{A}$, however we are able to measure it. It could naturally be included in any reasonable regression modeling scheme. Intuition from causal variable selection techniques would tell us to include $\boldsymbol{BAV}$ in the regression to reduce bias because it forms a confounding path ($\boldsymbol{A} \leftarrow \boldsymbol{BAV} \rightarrow \boldsymbol{Y}$). However, as has been demonstrated \citep{pearl2012class,pearl2011invited,ding2017instrumental,middleton2016bias} blocking this confounding path can actually increase or amplify the bias relative to the naive estimate only including $\boldsymbol{A}$.\\
\indent Here we will consider a special case with a linear system of equations. The target estimand is the average causal effect (ACE), which is simply $\beta_a$ in the linear model case (See appendix section \ref{subsc:derivations}). $\boldsymbol{U}$ is unmeasured and thus we cannot identify the ACE from the observed data, but we are interested in estimating the quantity with as little bias as possible.\\
\indent The true model representing Figure \ref{fig:DAG1} under the linear association assumption are:

\begin{align}
\boldsymbol{Y} &= \alpha_y + \boldsymbol{A}\beta_a + \boldsymbol{U}\beta_u +\boldsymbol{BAV}\boldsymbol{\beta_{bav}} + \boldsymbol{\epsilon_1}\label{Y truth}\\
\boldsymbol{A} &= \alpha_a + \boldsymbol{U}\gamma_u + \boldsymbol{BAV}\boldsymbol{\gamma_{bav}} + \boldsymbol{\epsilon_2}\label{A truth}
\end{align}
where $\alpha_y$ and $\alpha_a$ are the intercept terms for Y and A respectively. We use the form $\beta_x$ throughout this paper to denote true linear regression coefficients for some variable $X$ on the outcome Y. For example, the true regression coefficient for U on Y is $\beta_u$. Analogously, the true regression parameter for some variable $\boldsymbol{X}$ on the treatment A is represented by $\gamma_x$. The estimates of these parameters by OLS are denoted by $\hat{\beta_x}$, $\hat{\gamma_x}$ with additional superscripts to clarify which set of estimating equations the estimator is derived from. By assumption $\boldsymbol{\epsilon_1}$ and $\boldsymbol{\epsilon_2}$ are error terms independent of each other and all other variables represented in the DAG. We assume that $\boldsymbol{\epsilon_1}$ and $\boldsymbol{\epsilon_2}$ have mean 0, and some variance $\sigma^2_{\epsilon_{1,2}}$. In simulation experiments, we additionally assume that the error terms are normally distributed, but this is more than what is necessary for the theoretical results to hold.

\subsection{Matrix Notation and Probability Limits}

To tackle the question of model selection we must derive properties of the feasible $\hat{\beta_a}$ estimators. To this aim we propose expressing OLS estimates using matrix notation and ideas borrowed from the partial regression literature. Further, we propose considering also the probability limits of the estimators to extend our results to more general and realistic cases of bias amplification (See appendix \ref{sbsec:convergence in probability}). For the naive estimator we are estimating the following simple regression:

\begin{align}
\boldsymbol{Y} = \alpha_y + \beta_a\boldsymbol{A} + \boldsymbol{\upsilon_1}\label{seq naive}
\end{align}

Notice that $\boldsymbol{\upsilon_1}$ represents the error term in the estimating equation as opposed to $\boldsymbol{\epsilon_1}$ in the true underlying model. We can write $\boldsymbol{\upsilon_1}$ as , $\boldsymbol{\upsilon_1} = \boldsymbol{U}\beta_u +\boldsymbol{BAV}\boldsymbol{\beta_{bav}} + \boldsymbol{\epsilon_1}$. Unbiased estimation of the the naive model by OLS requires the assumption that $E[\boldsymbol{\upsilon_1}|\boldsymbol{A}] = 0$, but this of course is not true. The bias is a result of this erroneous assumption. The naive estimator bias is a special case of the classic omitted variables problem, where we have two omitted variables which are related to both the treatment and the exposure, $\boldsymbol{U}$ and $\boldsymbol{BAV}$.\\
\indent Let $\hat{\beta_a}^{naive}$ be the estimate of $\beta_a$ from the naive model \eqref{seq naive}. Throughout this paper we will consider the matrix $\boldsymbol{Z}$ to be a matrix of all the variables that we include in a regression that are not the variable of interest $\boldsymbol{A}$, in other words control variables in a selection on observables approach. In the naive model, $\boldsymbol{Z} = \boldsymbol{1}$ where throughout $\boldsymbol{1}$ will denote an $n\times 1$ vector of 1s. In matrix notation, applying the Frisch-Waugh-Lovell (FWL) theorem (see appendix \ref{sbsec: matrix and fwl}), we can write $\hat{\beta_a}^{naive}$ as:

\begin{align}
\hat{\beta_a}^{naive} &= \frac{\boldsymbol{A^TM_{\boldsymbol{1}}Y}}{\boldsymbol{A^TM_{\boldsymbol{1}}A}}\\\label{beta naive 1}
\end{align}
where $\boldsymbol{M_{\boldsymbol{1}}}$ is a centering projection matrix, defined and described in detail in appendix section \ref{sbsec: matrix and fwl}. In the case of linear relationships between all the variables,  following \cite{pearl2012class}, this estimator has the following expectation:

\begin{align}
E[\hat{\beta_a}^{naive}] = \beta_a + \beta_u\frac{\gamma_u\sigma_u^2}{\sigma_a^2} + \beta_{bav}\frac{\gamma_{bav}\sigma_{bav}^2}{\sigma_a^2}\label{exp: naive pearl}
\end{align}

The absolute bias for the ACE then clearly is $|\beta_u\frac{\gamma_u\sigma_u^2}{\sigma_a^2} + \beta_{bav}\frac{\gamma_{bav}\sigma_{bav}^2}{\sigma_a^2}|$. Now consider the estimates resulting from further conditioning on the observable $\boldsymbol{BAV}$ variable, i.e fitting the following model:

\begin{align}
\boldsymbol{Y} = \alpha_y + \boldsymbol{A}\beta_a + \boldsymbol{BAV}\beta_{bav} + \boldsymbol{\upsilon_2}\label{seq iv}
\end{align}

We will denote the resulting estimator $\hat{\beta_a^{|bav}}$ which can be written as follows by again applying the FWL theorem:
\begin{align}
 \hat{\beta_a^{|bav}}  = \boldsymbol{\frac{A^TM_zA}{A^TM_zY}} \label{conditional estimate matrix form}
\end{align}
where $\boldsymbol{Z} = \boldsymbol{[\one,BAV]}$ and $\boldsymbol{M_z}$ is the annihilator projection matrix of the matrix Z (see appendix \ref{sbsec:appendix-matrix} for details and properties). Again following Pearl \cite{pearl2012class}, the expectation of $\hat{\beta_a^{|bav}}$ is:

\begin{align}
E[\hat{\beta_a^{|bav}}] = \beta_a + \beta_u\frac{\gamma_u\sigma_u^2}{\sigma_a^2 - \gamma_{bav}^2\sigma_{bav}^2}\label{exp pearl bav}
\end{align}

In the appendix (see \ref{pearl derivation}) we explicitly show Pearl's derivation and how it relies on the conditional expectation $E[\boldsymbol{U}|\boldsymbol{A}, \boldsymbol{BAV}]$ being linear in both $\boldsymbol{A}$ and $\boldsymbol{BAV}$. Pearl's derivation is limited in that it is cumbersome and does not generalize well to a broad class of DAGs and functional forms. A simple example where we are unable to use Pearl's method is the case of an interaction term in the exposure structural equation between $\boldsymbol{U}$ and $\boldsymbol{BAV}$. Suppose we replace  equation \eqref{A truth} with:

\begin{align}
    \boldsymbol{A} &= \alpha_a + \boldsymbol{U}\gamma_u + \boldsymbol{BAV}\gbav +  \boldsymbol{U}\times \boldsymbol{BAV}\gamma_{u\times bav} + \boldsymbol{\epsilon_2}\\
    \implies \boldsymbol{U} &= \frac{\boldsymbol{A} - \alpha_a - \boldsymbol{BAV}\gamma_{bav} - \epsilon_2}{\boldsymbol{BAV}\gamma_{u\times bav} - \gamma_u}
\end{align}

The above equations show that $E[\boldsymbol{U}|\boldsymbol{A},\boldsymbol{BAV}]$ is nonlinear in $\boldsymbol{A}$ in $\boldsymbol{BAV}$, and cannot be represented by an unbiased least squares projection of the form $\boldsymbol{U} = \alpha_u + \boldsymbol{A}\zeta_a + \boldsymbol{BAV}\zeta_{bav} + \boldsymbol{\epsilon_3}$ as required by Pearl's derivation method (see Appendix \ref{pearl derivation} for details), where $\zeta_i$ represents the true regression coefficient for variable $i$. If we impose further strict distributional assumptions over all the variables, we may still be able to directly solve the conditional expectation and find an expression for bias in terms of the underlying parameters. In many applied cases, these distributional assumptions will not be justified, particularly assuming a distribution for the unmeasured confounding which will always be untestable.\\ 
\indent In contrast, if we consider the probability limits, we do not need to assume that $E[\boldsymbol{U}|\boldsymbol{A}, \boldsymbol{BAV}]$ is linear, nor do we have to make any additional distributional assumptions to find meaningful limiting expressions for our estimators in a broad class of clinically relevant circumstances. In addition to giving rise to a meaningful interpretation, the closed form asymptotics we derive allow us to more easily harness domain knowledge about the underlying causal process for the purpose of model selection.\\
\indent Since we are still  interested in the finite sample expectation of the estimators and the bias directly, we report the expectations when appropriate and feasible. The probability limit facilitates insight under weaker assumptions than those necessary to derive exact forms of the expectations. Additionally, in some cases, like the linear model of Pearl \citep{pearl2012class}, the probability limits for $\hat{\beta_a}^{naive}$ and $\hat{\beta_a^{|bav}}$ are precisely equal to their expectations (see appendix \ref{sbsec: plim calculations}).\\

\section{Treatment Variance as the root of bias amplification} \label{sc:variance accounted for}

Pearl \cite{pearl2012class} presented bias amplification results under the assumption of standard normal variables. Here we do not make any assumptions about distribution, mean, or variances for two reasons. First,  these assumptions are not strictly necessary to the result. Second, avoiding these assumptions helps  clarify some of the mechanics and the intuition behind the phenomenon of bias amplification. As the amplifying term in the denominator of equation \eqref{exp pearl bav}, $\sigma_a^2 - \gamma_{bav}^2\sigma_{bav}^2$, gets smaller the bias due to the unmeasured confounding path ($\boldsymbol{A} \leftarrow \boldsymbol{BAV} \rightarrow \boldsymbol{Y}$), $\beta_u\times\gamma_u\sigma_u^2$, increases. This is because when we specify the functional form of a system of random variables and conditional independence assumptions, we are also determining a formula for its variance. Under 1) the structural equation we specified for the exposure (equation \eqref{A truth}), and 2) the independence assumption between the unmeasured confounding and the bias amplifying variables, the variance is equivalent to:

\begin{align}
\sigma_a^2 &= \gamma_u^2\sigma_u^2 + \gamma_{bav}^2\sigma_{bav}^2 +\sigma_{\epsilon_2}^2\label{eq:variance determination}
\end{align}

Rearranging equation (15) to $\sigma_a^2 - \gamma_{bav}^2\sigma_{bav} = \gamma_u^2\sigma_u^2  + \sigma_{\epsilon_2}^2$, it becomes clear  the residual variance in $\boldsymbol{A}$ (i.e. not due to the $\boldsymbol{BAV}$ variable) is equal to the sum of the variance due to $\boldsymbol{U}$ and the independent variation $\sigma_{\epsilon_2}$. Therefore, the amplification of the bias in the general case depends not only on the magnitude of $\gamma_{bav}$ (i.e the strength of association between $\boldsymbol{BAV}$ and $\boldsymbol{A}$), but how much of the treatment variance the $\boldsymbol{BAV}$ variable linearly accounts for. When we assume that all the variables are standard normal, the amplification becomes $1 -\gamma_{bav}^2$ as presented in Pearl\cite{pearl2012class},because the variance of standard normal variables is equal to 1 ($\sigma_a^2,\sigma_{bav}^2 = 1$).\\
\indent In order to visualize this phenomenon, we use ideas from partial regression plots \citep{velleman1981efficient}. By the FWL theorem, we can always pre-multiply an estimating equation by the residual-making variables of a set of regressors and get the same estimates (see appendix \ref{sbsec: matrix and fwl} for further details). For example, the following two regression equations produce the same numerical estimates of $\hat{\beta_a}^{naive}$:

\begin{align}
\boldsymbol{Y} &= \alpha_y + \boldsymbol{A}\beta_a + \boldsymbol{\upsilon_1}\label{regression1}\\
\boldsymbol{M_{\boldsymbol{1}}Y} &= \boldsymbol{M_{\boldsymbol{1}}A}\beta_a + \boldsymbol{\upsilon_1}\label{eq:modified}
\end{align}

Equation \eqref{eq:modified} is the model for a simple linear regression of a modified outcome, $\boldsymbol{M_{\boldsymbol{1}}Y}$ on a modified treatment, $\boldsymbol{M_{\boldsymbol{1}}A}$ (See appendix \ref{sbsec: matrix and fwl}). There is no intercept term as the mean of the modified treatment
must be equal to zero. 

\begin{align}
\boldsymbol{Y} &= \alpha_y + \boldsymbol{A}\beta_a + \boldsymbol{BAV}\beta_{bav} + \boldsymbol{\upsilon_2}\label{eq:cond2}\\ 
\boldsymbol{M_zY} &= \boldsymbol{M_zA}\beta_a + \boldsymbol{\upsilon_2}\label{eq:cond modified}
\end{align}

Similarly equations \eqref{eq:cond2} and \eqref{eq:cond modified} produce equivalent estimates of $\hat{\beta}_a$, where $\boldsymbol{Z} = [\boldsymbol{1} \quad \boldsymbol{BAV}]$ is a column of 1s and the $\boldsymbol{BAV}$ variable. Equation \eqref{eq:cond modified} is a single variable regression on a transformed set of variables. The modified $\boldsymbol{Y}$ is produced by taking the residuals from regressing $\boldsymbol{Y}$ on a column of 1s and $\boldsymbol{BAV}$, in other words the dependent variable is the remaining variation in $\boldsymbol{Y}$ which is not linearly associated with an intercept and $\boldsymbol{BAV}$. The independent variable is the remaining variation in $\boldsymbol{A}$ not linearly associated with a column of 1s and $\boldsymbol{BAV}$. Since we have reduced the multi-variable regression to a simple linear regression we can easily visualize the amplification process via a partial regression plot.
\begin{figure}[H]
	\centering
\includegraphics[scale=.5]{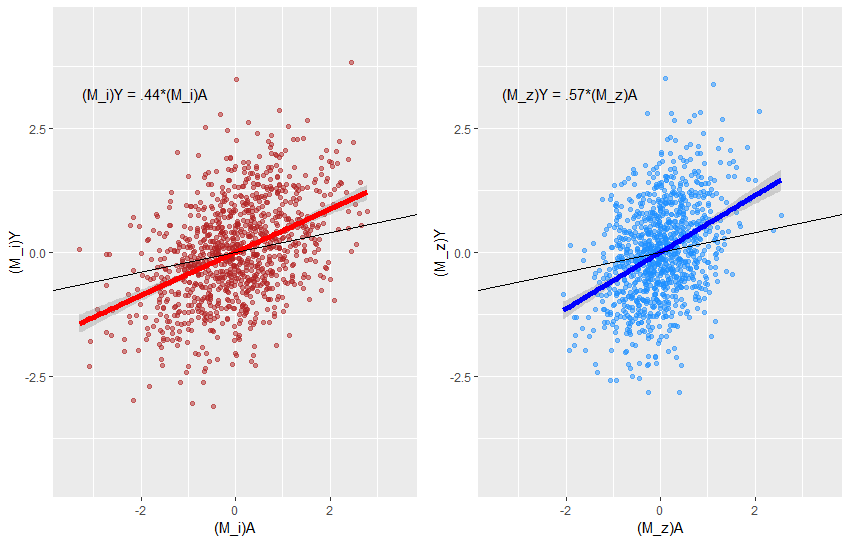}
\caption{In both panels, the unbiased ACE ($\beta_a = 0.2$) is shown by the dotted black line. In the left panel, the red dots represent the centered treatment ($\boldsymbol{M_\iota A}$) plotted against the centered outcome ($\boldsymbol{M_\iota Y}$) and the estimated slope $\naive$ is shown with the bolded red line. This represents the equivalent regressions in equations \eqref{regression1} and \eqref{eq:modified}. In the right panel the blue dots represent the modified treatment ($\boldsymbol{M_zA}$) plotted against the modified outcome ($\boldsymbol{M_zY}$). The solid blue line represents the treatment estimate from the equivalent regressions \eqref{eq:cond2} and \eqref{eq:cond modified}.}
\label{biasviz}
\end{figure}

In Figure \ref{biasviz} the left visualizes the naive regression equation \eqref{eq:modified}, whereas the blue graph on the right visualizes the regression equation \eqref{eq:cond modified} that includes BAV. The data was simulated from a special case of equations \eqref{Y truth} and \eqref{A truth}, with $n= 1000$. Details can be found in the appendix (Section \ref{sbsec:simfig}).\\
\indent The unbiased ACE is the slope of the black line ($\beta_a = 0.2$) in these plots. The slope of the blue line (equal to the OLS estimator from the amplifying model) is clearly farther away from the true slope (in black) compared to the slope of the red line from the naive model, and thus the conditional estimator is more biased.\\
\indent Note first that including $\boldsymbol{BAV}$ in the model reduces the variance in the adjusted treatment, which can be seen by comparing the relative sparsity of points along the x-axis in red compared to the relative density  of points along the x-axis in blue. However, if we inspect the spread of points vertically along the y-axis, we can see that the red and blue samples are similarly dispersed in this dimension because conditional on the treatment, linear combinations of $\boldsymbol{BAV}$ explain very little of the variance in the outcome. Most importantly, including $\boldsymbol{BAV}$ does not change the variance in $\boldsymbol{Y}$ due to $\boldsymbol{U}$, the unmeasured confounder. As a result, the line of best fit of the adjusted model must be steeper in absolute terms in order to maintain the association between the adjusted response and adjusted treatment over the narrower variation of the adjusted treatment variable. When we add the $\boldsymbol{BAV}$ to the regression model, the bias is 0.14 larger in absolute terms (or approximately 65\% greater in relative terms) than the naive estimate, even though it blocks a confounding path between the treatment $\boldsymbol{A}$ and the outcome $\boldsymbol{Y}$. More simply, trying to block a confounding path with weak response association can amplify bias in causal effect estimation because it increases the proportion of treatment association due to unmeasured confounding on unblocked paths.\\
\indent The magnitude of bias amplification can be potentially very large. The absolute bias of the $\boldsymbol{BAV}$ estimator will be larger than the absolute bias of the naive estimator whenever $|\beta_u\frac{\gamma_u\sigma_u^2}{\sigma_a^2} + \beta_{bav}\frac{\gamma_{bav}\sigma_{bav}^2}{\sigma_a^2}| < |\beta_u\frac{\gamma_u\sigma_u^2}{\sigma_a^2 - \gamma_{bav}^2\sigma_{bav}^{bav}}|$ if the relationships are linear. In particular, the bias is greater if $\boldsymbol{BAV}$ is not strongly associated with the outcome (i.e small values of $\beta_{bav}\sigma_{bav}^2$)). The special case $\beta_{bav} = 0$  implies that $\boldsymbol{BAV}$ is a true instrumental variable. Instrumental variables were in fact the leading case for the discovery of this class of bias amplifiers\cite{WOOLDRIDGE2016232,pearl2012class}. If there are no interaction terms (i.e a model that is linear in the original variables) adding an instrumental variable always weakly increases absolute bias in OLS relative to the naive model, with equality only when there is no unmeasured confounding \citep{WOOLDRIDGE2016232,pearl2012class}.\\
 \indent In summary, variable selection approaches which aggressively target confounding paths with strong associations with treatment and weak associations with outcome are at grave risk of bias amplification as they are much more sensitive to the assumption that a full sufficient set is measurable. Adding controlling variables in proportion to their ability to predict the treatment in linear models only becomes a bias reducing approach if the resulting variable set satisfies ignorability assumptions. This is often not possible or extremely unlikely in many non-experimental settings.

\section{Generalizing to a larger class of causal models}
\label{sec:identify_amp}

The danger of including variables that are strongly associated to the treatment is that we cannot identify unmeasured confounding. Consider the probability limit of the estimator in equation (\ref{conditional estimate matrix form}):

\begin{align}
\hat{\beta_a}^{|bav} &\overset{p}{\to} \frac{COV(\boldsymbol{A,U})}{\underset{n\to\infty}\plim\frac{1}{n}\sum_{i=1}^{n}\hat{\xi_i}^2}\label{eq:residual plim}
\end{align}

\noindent where $\sum_{i=1}^{n}\hat{\xi_i}^2$ is the estimated sum of squared residuals from the regression of the treatment on $\boldsymbol{BAV}$ and an intercept term (note that $\boldsymbol{\xi} = \boldsymbol{U}\gamma_u + \boldsymbol{\epsilon_2}$ from equation \eqref{A truth}). Since all the variables necessary to estimate $\hat{\xi}$ are observable, we can identify the denominator, or the amplifying term. Equation \eqref{eq:residual plim} shows that the amplifying term is numerically equivalent to the sum of squared residuals of $\boldsymbol{A}$ on an intercept column and $\boldsymbol{BAV}$. Note that this result does not even require a limit or expectation to hold. However, it is typically more useful to think about the probability limit in an applied application since the numerator simplifies to a covariance term in the case that the added variables are independent of the unmeasured confounding.\\
\indent In the case that individual treatment assignments are independent and have variance 1, the probability limit of the average of the squared residuals is equal to one minus the proportion of variance of $\boldsymbol{A}$ explained by $\boldsymbol{Z}$  \citep{middleton2016bias}, i.e. 

\begin{align}
 R^2_{\boldsymbol{A}|\boldsymbol{1},\boldsymbol{BAV}} &= 1 - \frac{\underset{n\to\infty}\plim \frac{1}{n}\boldsymbol{A^TM_zA}}{\underset{n\to\infty}\plim \frac{1}{n}\boldsymbol{A^T M_1 A}}\\
 \implies \underset{n\to\infty}\plim \frac{1}{n}\boldsymbol{A^TM_zA} &= (1- R^2_{\boldsymbol{A}|\boldsymbol{1},\boldsymbol{BAV}}) \\
 &  \text{ if } \underset{n\to\infty}\plim \frac{1}{n}\boldsymbol{A^T M_1 A} = 1. 
 	\end{align}
 
 Thus far we have only considered the DAG in Figure \ref{fig:DAG1} under the restrictive assumption of linear associations amongst variables. The identifiability of the bias amplification term of the preceding section can be extended in two important ways. First,  the results extend to the addition of any $p$ bias-amplifying variables by simply increasing the number of columns of $\boldsymbol{Z}$ to include any number of bias-amplifying variables, as the FWL theorem allows for arbitrary numbers of columns as long as $\boldsymbol{Z}$ is of full rank. We include examples of multiple bias-amplifying variables in section \ref{sc:real data}. Second,  in the following subsection, we provide the details of how to extend the result to non-linear associations.

\subsection{Non-Linear Associations}
\label{sbsec:non-linear}
In the previous sections we assume that the data generating processes governing the treatment and the outcome are linear. However, in order to identify the amplification term with observable variables this is not strictly necessary. First, we relax the assumption that the data generating process of $\boldsymbol{A}$ is linear, allowing it to be any arbitrary function $\boldsymbol{A} = f(\boldsymbol{U},\boldsymbol{BAV}, \boldsymbol{\epsilon_2})$ but let the model for $\boldsymbol{Y}$ remain unchanged such that equation \eqref{Y truth} holds. In the appendix (section \ref{sbsec: plim calculations}) we show that the numerical form of the conditional estimator to be $\bav = \beta_a + \beta_u\boldsymbol{\frac{A^TM_zU}{A^TM_z}} + \boldsymbol{\frac{A^TM_z\epsilon_2}{A^TM_zA}}$, where $\boldsymbol{M_zA}$ is by definition the vector of residuals from the regression of $\boldsymbol{A}$ on the columns of $\boldsymbol{Z}$, which we specified to mean $\boldsymbol{BAV}$ and a column of 1s. As in the fully linear case, $\boldsymbol{A^TM_zA}$ is the sum of squared residuals. The residuals will be the treatment $\boldsymbol{A}$, removed of linear components of $\boldsymbol{Z}$ which do not directly depend on the underlying function governing the relationship between $\boldsymbol{A}$ and $\boldsymbol{BAV}$. Note that $\boldsymbol{A^TM_zA}  \leq \boldsymbol{A^TM_{\boldsymbol{1}}A}$ since $\boldsymbol{Z}$ contains the column of 1's, and so amplification will occur as long as the treatment is some function of $\boldsymbol{BAV}$ producing a positive correlation between the treatment and bias amplifying variable. As shown in section \ref{sc:variance accounted for}, the extent of the amplification will be determined by the linear correlation between $\boldsymbol{A}$ and $f(\boldsymbol{U},\boldsymbol{BAV},\boldsymbol{\epsilon_2})$.\\
\indent When we allow for non-linear associations in the outcome, an important point to clarify is that adjusting for $\boldsymbol{BAV}$ in OLS will not necessarily be sufficient to block the confounding path that $\boldsymbol{BAV}$ forms. Consider a very simple extension to the outcome model as follows:

\begin{align}
\boldsymbol{Y} &= \alpha_y + \boldsymbol{A}\beta_a +\boldsymbol{U}\beta_u + \boldsymbol{BAV}\beta_{bav} + \boldsymbol{BAV^2}\beta_{bav^2} + \boldsymbol{\epsilon_1}\label{eq:Y nonlinear}\\
\boldsymbol{A} &= \alpha_a + \boldsymbol{U}\gamma_u + f(\boldsymbol{BAV})\gamma_{bav} + \boldsymbol{\epsilon_2}\label{eq:A nonlinear}
\end{align}

\noindent where for simplicity $f(\boldsymbol{BAV})$ is a high-ordered polynomial term that is correlated with $\boldsymbol{BAV^2}$ after adjusting for the linear term. If we adjust for $\boldsymbol{BAV}$ and not the squared term, the causal estimates will clearly suffer from omitted variable bias since the squared term remains correlated with both the treatment and outcome.  However, we can still identify the amplification. The proper ACE under equation (\ref{eq:Y nonlinear}) are $\frac{\partial E[\boldsymbol{Y}|\boldsymbol{A},\boldsymbol{U},\boldsymbol{BAV}]}{\partial \boldsymbol{A}} = f_1^{\prime}(\boldsymbol{A})\beta_a$. Bias needs to be evaluated as deviations from the true causal effects ($f_1^{\prime}(\boldsymbol{A})\beta_a$) and not the parameter $\beta_a$.

\begin{align}
\hat{\beta_a^{|bav}} &\overset{p}{\to} \beta_a + \beta_u\frac{COV(\boldsymbol{A,U})}{\frac{1}{n}\sum_{i=1}^n \hat{\xi_i}^2} + \beta_{bav^2}\frac{COV(\boldsymbol{A,BAV^2})}{ \frac{1}{n}\sum_{i=1}^n \hat{\xi_i}^2} \label{plim: nonlinear cond}
\end{align}

\noindent where $\hat{\xi_i}^2$ are the squared residuals from the regression of $\boldsymbol{A}$ on $\boldsymbol{BAV}$ and a constant. Although the amplified bias is different from the simple case, the factor by which the bias is amplified is still identifiable using a regression depending only on observables. The amplification results from misspecification of the relationship between $\boldsymbol{BAV}$ and the outcome. Similar to before there will be some cases where the amplification of the U-bias is outweighed by the reduction in omitted variable bias due to $\boldsymbol{BAV}$ and other cases in which this would not be the case.\\
\indent Now consider a fully non-linear, but still additive, model specification:

\begin{align}
\boldsymbol{Y} &= \alpha_y + f_1(\boldsymbol{A})\beta_a + f_2(\boldsymbol{U})\beta_u + f_3(\boldsymbol{BAV})\beta_{bav} +\boldsymbol{\epsilon_1}\\
\boldsymbol{A} &= \alpha_y + g_1(\boldsymbol{U})\gamma_u + g_2(\boldsymbol{BAV})\gbav + \boldsymbol{\epsilon_2}
\end{align}
If we estimate the linear naive and linear adjusted models as before we get the following estimates:

\begin{align}
\naive &\overset{p}{\to} \beta_a\frac{COV(\boldsymbol{A},f_1(\boldsymbol{A}))}{\siga}+ \beta_u\frac{COV(\boldsymbol{A},f_2(\boldsymbol{U}))}{\siga} +\beta_{bav}\frac{COV(\boldsymbol{A},f_3(\boldsymbol{BAV}))}{\siga}\label{plim:naive nonlinear}\\
 \hat{\beta_a}^{|bav} &\overset{p}{\to} \beta_a\frac{\plim \frac{1}{n}\boldsymbol{A^TM_z} f_1(\boldsymbol{A})}{\plim \frac{1}{n} \sum_{i=1}^n \hat{\xi_i}^2}+ \beta_u\frac{COV(\boldsymbol{A},f_2(\boldsymbol{U}))}{\plim\frac{1}{n} \sum_{i=1}^n \hat{\xi_i}^2} + \beta_{bav}\frac{\plim \frac{1}{n}\boldsymbol{A^TM_z} f_3(\boldsymbol{BAV})}{\plim \frac{1}{n} \sum_{i=1}^n \hat{\xi_i}^2}\label{plim:cond nonlinear}
\end{align}

Looking at equation \eqref{plim:cond nonlinear}, the unmeasured confounding pathway remains amplified and we can estimate the residuals which cause the amplification from observable quantities. However, the direction of the shift in overall bias is unclear when we condition on $\boldsymbol{BAV}$. Note that the middle term is unambiguaously larger for the $\boldsymbol{BAV}$ model than the naive model. The third term in the $\boldsymbol{BAV}$ model will be smaller than the naive model in the numerator, but larger in the denominator. Most troubling, by allowing $\boldsymbol{Y}$ to be a nonlinear function of $\boldsymbol{A}$, we can no longer predict whether the first term is getting closer or farther from the truth. However, if $f_1(\boldsymbol{A})$ and $f_3(\boldsymbol{BAV})$ are known or can be well approximated, we can estimate $\frac{\plim \frac{1}{n}\boldsymbol{A^TM_z} f_1(\boldsymbol{A})}{\plim \frac{1}{n} \sum_{i=1}^n \hat{\xi_i}^2}$ and $\frac{\plim \frac{1}{n}\boldsymbol{A^TM_z} f_3(\boldsymbol{BAV})}{\plim \frac{1}{n} \sum_{i=1}^n \hat{\xi_i}^2}$, since they are functions entirely of observables. To estimate the numerators, three regressions should be run: the exposure on $\boldsymbol{Z}$, $f_1(\boldsymbol{A})$ on $\boldsymbol{Z}$, and $f_3(\boldsymbol{BAV})$ on $\boldsymbol{Z}$. By storing the residuals and combining them appropriately, the first and third term in equation \eqref{plim: nonlinear cond} can be estimated up to $\beta_a$ and $\beta_{bav}$ (For more details see appendix \ref{sbsec: matrix and fwl}).\\
\indent If we make no assumptions about the functional form, and allow for non-linearities, and interactions between all variables, we can show that the OLS adjusted estimator is always the expression below (see appendix \ref{sbsec: matrix and fwl}):
\begin{align}
\hat{\beta_a^{|z}} &= \frac{\boldsymbol{(M_zA)^TM_zY}}{\boldsymbol{(M_zA)^T(M_zA)}}
\end{align}

When $\boldsymbol{Z}$ includes an intercept column, both $\boldsymbol{(M_zA)^T}$ and $\boldsymbol{M_zY}$ will have mean zero and thus we can think of the numerator as an empirical estimate of the covariance between the residuals from the regression of $\boldsymbol{A}$ on $\boldsymbol{Z}$ and the residuals from the regression of $\boldsymbol{Y}$ on $\boldsymbol{Z}$. Unmeasured confounding bias in OLS occurs when after projecting out linear combinations of the controlling variables, $\boldsymbol{Z}$, there remain linear associations between the outcome and the treatment due to unobserved variables. The part of the bias due to unmeasured confounding is amplified whenever the control variables explain variance in the treatment. Holding all else constant, as the residuals from the regression of the treatment on $\boldsymbol{Z}$ decrease in magnitude, the absolute value of the estimator $\hat{\beta_a^{|z}}$ will increase in magnitude. This is a general form of the result we showed in the previous section which is extremely powerful in that it encaptures a very large class of structural equations and DAGs. However, the cost of this generality is that without making more specific assumptions about the particular form of the model, and in particular the outcome model, it becomes more difficult to incorporate the knowledge of the amplification factor into our model selection and thus apriori know which of the two estimators, $\naive$ or $\bav$, will be less biased. Interaction terms, for example, are an additional difficulty. Pearl \cite{pearl2012class}, for example, showed that under a simple interaction effect between the unmeasured confounding and some function of a pure instrument, the adjusted estimator can be less biased than the naive case. To properly evaluate estimators in the context of bias amplification requires appropriate simulations. In the next section, we describe how to avoid the pitfalls of previous simulation work \cite{meyers2011}.\\

\section{Causal Simulation Experiments: The Case of  Bias Amplification}\label{sc: new simulation}

In our experience, simulating bias amplification is challenging in a number of subtle, but important ways. Our context of interest is assessing the potential for bias amplification in an analysis of an observational study in which we have measured several independent variables and the outcome but there might be an unmeasured confounder. We are interested in evaluating the feasible estimators we have developed in the previous sections, $\naive$ and $\bav$ for example, with respect to possible data sets generated by a class of DAGs and structural equations. In this section we show that if we constrain certain aspects of the simulated data (in particular, the marginal variances of observed quantities), we are better able to articulate and answer causal questions about the effect of bias amplification on proposed estimators. While we discuss the example of bias amplification simulations specifically, this section has implications for simulating data to test causal estimators more broadly.

\begin{figure}[H]
\centering
\begin{subfigure}[]{.32\textwidth}
\includegraphics[height = 0.16\textheight]{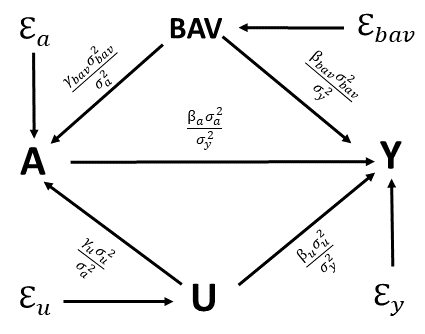}
\caption{}
\label{ext: 0 squiggle}
\end{subfigure}
\begin{subfigure}{.32\textwidth}
\includegraphics[height = 0.16\textheight]{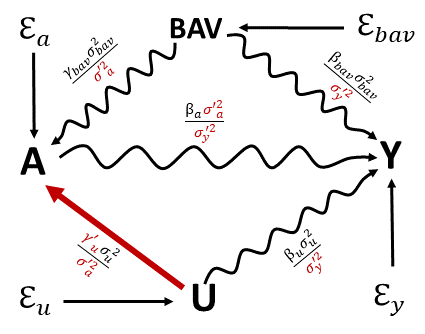}
\caption{}
\label{extended dag 4 squiggle}
\end{subfigure}
\begin{subfigure}{.32\textwidth}
\includegraphics[height = 0.16\textheight]{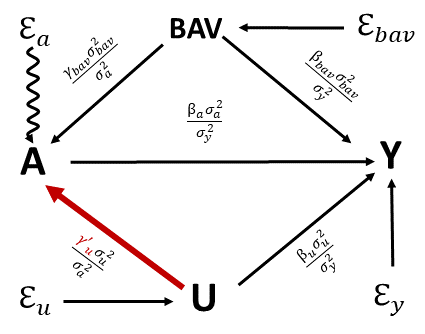}
\caption{}
\label{extended 1 squiggle}
\end{subfigure}
\caption{Extended causal diagrams \cite{robins2010alternative} where (a) represents the underlying causal structure. In addition to the usual causal pathways we explicity show the pathway of the independent error terms, $\epsilon_a, \epsilon_u, \epsilon_{bav}, \epsilon_y$. (b) shows the simulation experiment where we change strength of the edge $\boldsymbol{U} \to \boldsymbol{A}$ (shown in bold red) and do not renormalize the variance of the treatment ($\boldsymbol{A}$). All edges which have been inadvertently modified are symbolized as squiggly arrows. All parameters which have been modified (inadvertently or intentionally) are shown in red along the edges. c) Intervening on $\boldsymbol{U}\to \boldsymbol{A}$, where the variance of the treatment is fixed. To do so we modify the variance of the noise term $\epsilon_a$, visualized by the squiggly arrow. Notice no other edges are inadvertently modified.}
\end{figure}
 
Now, consider the challenge of determining the effect of increasing unmeasured confounding on bias amplification in Figure \ref{ext: 0 squiggle}. We might, for example, be interested in how large an unmeasured confounder must be, with fixed amplifying variables, to cross some threshold of bias in the adjusted model as part of a sensitivity analysis. To answer such a question we must define clearly what is meant by the strength of an unmeasured confounder. In Figure \ref{ext: 0 squiggle}, there are two edges which determine the overall bias due to the unmeasured confounding path through U: the edge from $\boldsymbol{U}$ to $\boldsymbol{A}$ and the edge from $\boldsymbol{U}$ to $\boldsymbol{Y}$. The bias due to the unmeasured confounding path through U in the naive model is simply the product of the weight of these two edges, scaled by the variance of the treatment as shown in equation \eqref{beta naive 1}. The extent to which bias can become amplified, however, is not symmetric with respect to the weight of the edges $\boldsymbol{U} \rightarrow \boldsymbol{A}$ and $\boldsymbol{U} \rightarrow \boldsymbol{Y}$, since amplification is the result of variance explained in the treatment as discussed in section \ref{sc:variance accounted for}. There is more potential for amplification of a strong unmeasured confounder (in the sense the product of the confounding edges is large) when the strength is due to $\boldsymbol{U}$ being a strong cause of $\boldsymbol{Y}$ compared to a strong cause of $\boldsymbol{A}$. This is because when $\boldsymbol{U}$ is a strong cause of $\boldsymbol{A}$, the $\boldsymbol{BAV}$ can only explain a small amount of the variance of $\boldsymbol{A}$, limiting the possible amount of bias amplification. Thus to answer a causal question about the effect of increased unmeasured confounding on bias amplification we should only vary one of the confounding edges, holding all other edges fixed.\\
\indent As an example, suppose we are interested in the change in bias amplification when we increase the strength of the edge from $\boldsymbol{U}$ to $\boldsymbol{A}$, holding all else constant. This notion of intervening on a single edge of our DAG while holding the others fixed should be familiar to causal inference practitioners since it is the principle behind counterfactual analysis more broadly. Here we want to ensure that our results from varying a single edge are not confounded by variations in other edges as the result of unintended consequences or induced associations.\\  
\indent Because the goal is to increase the strength of a single edge, holding all else constant, we must specify a metric by which we measure the strength of the edge. In a fully linear system, we might consider the strength of the edge as the regression coefficient itself, $\gamma_u$, or the proportion of variance explained by $\boldsymbol{U}$, $\frac{\gamma_u^2\sigma_u^2}{\sigma_a^2}$ and the sign of $\gamma_u$. It is tempting to see the two measures as equivalent with different scalings, but this is only true in the context of simulating a single equation. In the context of a system of linear equations, especially with the potential for bias amplification, we argue the relevant quantity is the proportion of variance explained by each child node of the parent variable. This can be seen most easily by examining the bias formula in equation \eqref{exp pearl bav}, where the amplifying term is the remaining variation in $\boldsymbol{A}$ unexplained by the potential bias amplifying variables.\\
\indent Consider the implications of treating the coefficients themselves as the relevant measure of edge strength in a simulation trying to determine the effect of increasing the causal association along the path from $\boldsymbol{U}$ to $\boldsymbol{A}$. If we want to increase $\gamma_u$ to $\gamma_u^\prime > \gamma_u$ without changing any other parameters, we must also increase the total variance in the treatment, $\boldsymbol{A}$, since $\sigma_a^2 = \gamma_u^2\sigma_u^2 + \gamma_{bav}^2\sigma_{bav}^2 + \sigma_{\epsilon_2}^2$. A treatment with a larger variance is in some sense a different intervention, and thus this simulation is not compatible with the class of experiments which generated the original data with parameter $\gamma_u$. Further, from the previous sections we know this implies the total amount of variance explained from the bias amplifier $\boldsymbol{BAV}$ is reduced, since $\frac{\gamma_{bav}^2\sigma^2_{bav}}{(\sigma_a^2)^\prime}$ has been reduced. Although we have not changed the parameter $\gamma_{bav}$ we have decreased the extent to which $\boldsymbol{BAV}$ amplifies the bias as seen by examining equation \eqref{exp pearl bav}. The increased variance in $\boldsymbol{A}$ in turn modifies the total variance of $\boldsymbol{Y}$. Therefore, the relative proportion of variance of $\boldsymbol{Y}$ that is explained by $\boldsymbol{BAV}$ is modified by changing the causal effect of $\boldsymbol{U}\rightarrow \boldsymbol{A}$, as are the measured proportion of variance of $\boldsymbol{BAV}\rightarrow \boldsymbol{A}$, $\boldsymbol{A} \rightarrow \boldsymbol{Y}$, $\boldsymbol{U}\rightarrow \boldsymbol{Y}$, and $\boldsymbol{BAV}\rightarrow \boldsymbol{Y}$ and their associated covariance terms.\\
\indent We can see in Figure \eqref{extended dag 4 squiggle} that by modifying a single coefficient and leaving all other coefficients unchanged we have inadvertently modified the relative proportion of variance explained by the 4 other edges ($\boldsymbol{BAV} \to \boldsymbol{A}$, $\boldsymbol{BAV} \to \boldsymbol{Y}$, $\boldsymbol{U} \to \boldsymbol{Y}$, and $\boldsymbol{A} \to \boldsymbol{Y}$) represented by the wavy arrows. Data generated by the second set of structural equations are not compatible with the constraints of the experiment which generated the first data and by intervening on a single edge we have modified all of the competing effects of interest. Comparing the distribution of estimates produced under $\gamma_u$ and $\gamma^{\prime}_u$ gives us a confounded and thus biased estimate of the impact of increasing the unmeasured confounding through its causal pathway to the treatment on the estimators or functions thereof. We will show that this bias can result in under-estimating the impact of bias amplifying variables.\\
\indent In general, when we vary one of the regression coefficients along a causal pathway, this has upstream and downstream effects on the proportion of variance explained by all variables going into or out of the varied node. In order to keep the proportional effects of the other edges constant, we need to use the error terms of the structural equations ($\boldsymbol{\epsilon_1}$ and $\boldsymbol{\epsilon_2}$) to absorb the shocks to the marginal variances.\\
\indent In Figure \ref{extended 1 squiggle}, if we change $\gamma_u$ and simply adjust the structural error term $\epsilon_a$ such that the total variance in $\boldsymbol{A}$ remains constant, we can isolate the effect of modifying $\boldsymbol{U}\to \boldsymbol{A}$. Below in Figure \ref{fig: raincloud 1} we visualize the consequences of failing to hold the variance of the treatment when we modify $\gamma_u$.

\begin{figure}[H]
\centering
\begin{subfigure}[c]{.44\textwidth}
\centering
\includegraphics[height = .2\textheight]{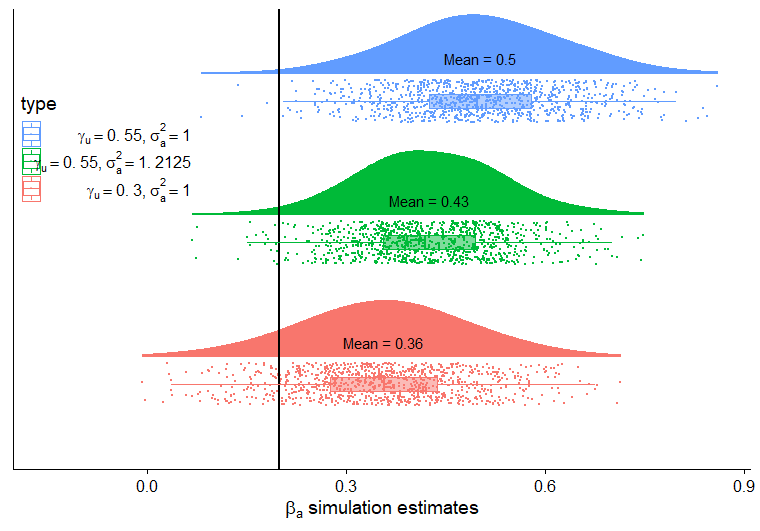}\\
\caption{}
\label{fig: raincloud 1}
\end{subfigure}
\begin{subfigure}[c]{.44\textwidth}
\centering
\includegraphics[height = .2\textheight]{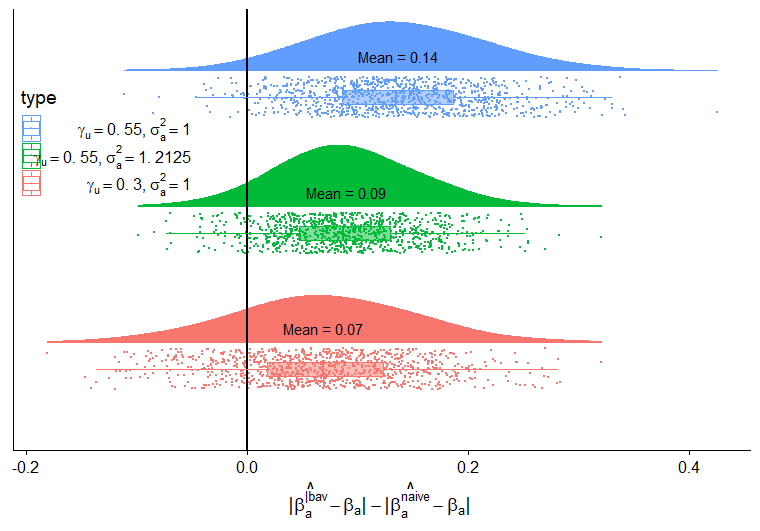}\\
\subcaption{}
\label{fig: raincloud 2}
\end{subfigure}
\caption{(a) Simulation results from the experiment of intervening on the edge $\boldsymbol{U} \to \boldsymbol{A}$. The ground truth, $\beta_a = 0.2$, is visualized by the black vertical line. In red we visualize the baseline bias amplification. Green shows the results of the conditional estimator in the case where we increase the weight of the edge, but fail to fix the variance of the treatment, allowing it to grow. In blue, we show the bias when we increase the weight of the edge but now hold the variance constant so the variances remain compatible with the original data. (b) Simulation results from the same experiment as (a) but the outcome is the difference in absolute bias between the conditional estimator $\bav$ and the naive estimator $\naive$. A value greater than zero (black vertical line) indicates that the conditional estimator was more biased than the naive estimator. Parameter values: $\beta_a = 0.2$, $\beta_u = 0.3$, $\beta_{bav} = -0.05$, $\gamma_{bav} = 0.6$}
\end{figure}

In red, for Figure \ref{fig: raincloud 1}, we simulate bias amplification where $\gamma_u = 0.3$. In green, we simulate bias amplfication where $\gamma_u$ is increased to $0.55$ holding all other parameters constant, thus allowing the total variance of the treatment to grow from $1$ to $1.21$. This has the downstream effect of also increasing the variance of the outcome from $1$ to $1.02$. This also then impacts the relative proportions of variance explained of the treatment and the outcome that are explained by $\boldsymbol{U}$ and $\boldsymbol{BAV}$ respectively. Notice that the bias increases from $20\%$ ($\frac{0.36-0.3}{0.3})$ to $43\%$ $(\frac{0.43-0.3}{0.3})$. In blue, we increase $\gamma_u$ from $0.3$ to $0.55$, but re-normalize the variance in the treatment to remain constant at $1$. The bias now increases further to $67\%$ ($\frac{0.5-0.3}{0.5}$) with respect to the original simulation in red. We do this by decreasing the variance of the independent noise term, $\boldsymbol{\epsilon_a}$ to $\boldsymbol{\epsilon^\prime_a}$, allowing it to absorb the increase in variation from $\boldsymbol{U}$. When we do not fix the variance, we underestimate the impact of the amplifier on both the bias and the variance {\em because the unfixed variance case simulates a different kind of intervention due to the change in variance of the treatment variable}. In the simulation above, by not keeping the variance fixed in $\boldsymbol{A}$ we implicitly reduced the amount of variance that $\boldsymbol{BAV}$ accounts for in the treatment from $36\%$ to $30\%$. In effect, we were comparing the distribution of

\begin{align*}
P(\hat{\beta_a}&|\gamma_u^\prime,\gamma_{bav}, \beta_u, \beta_a, \beta_{bav}) \\
\text{ to }\\
 P(\hat{\beta_a}&|\gamma_u,\gamma_{bav}, \beta_u, \beta_a, \beta_{bav})
\end{align*}

\noindent when a more fair causal counterfactual would be to compare the distribution of
\small
\begin{align*}
P(\hat{\beta_a}&|(\boldsymbol{U} \rightarrow \boldsymbol{A})^\prime, \boldsymbol{U} \rightarrow \boldsymbol{Y}, \boldsymbol{BAV} \rightarrow \boldsymbol{A}, \boldsymbol{BAV} \rightarrow \boldsymbol{Y}, \boldsymbol{A} \rightarrow \boldsymbol{Y})\\
\text{ to }\\
P\hat{\beta_a}&|\boldsymbol{U} \rightarrow \boldsymbol{A}, \boldsymbol{U} \rightarrow \boldsymbol{Y}, \boldsymbol{BAV} \rightarrow \boldsymbol{A}, \boldsymbol{BAV} \rightarrow \boldsymbol{Y}, \boldsymbol{A} \rightarrow \boldsymbol{Y}).
\end{align*}
\normalsize

Therefore, our simulation \textit{experiment} results in green are distorted because when we increased the unmeasured confounding through $\boldsymbol{U} \to \boldsymbol{A}$, we also decreased the strength of the bias amplifying variable through the pathway $\boldsymbol{BAV} \to \boldsymbol{A}$. Notice that this bias will impact decisions and conclusions we might make about the merits of different estimators in this context. For example, below we compare the conditional estimator, $\bav$, to the naive estimator, $\naive$ with respect to their bias in the same three simulation set ups.\\
\indent In Figure \ref{fig: raincloud 2} we show the direct comparison of the bias for the conditional and the naive estimators. When we increase the unmeasured confounding through $\gamma_u$ but fail to renormalize the treatment variance, we do not capture the full extent to which the conditional estimator amplifies the bias. If we compared the green and red plot it would seem that nearly doubling the unmeasured confounding coefficient only has a small impact on the relative bias of the naive and conditional estimator, since the relative bias only increased from 0.07 to 0.09 ($29\%$). By comparing the green density plot to the blue, we see the relative bias doubles (from 0.07 to 0.14). Therefore, the decision to use the naive or conditional estimator is in fact much more sensitive to the amount of unmeasured confounding than it would appear under the improper simulation with floating variance. It is extremely important to do these kinds of simulations properly particularly in the context of sensitivity analysis where we are testing the performance of estimators with respect to untestable assumptions such as unmeasured confounding.\\
\indent To properly simulate bias amplification and answer questions of clinical concern with respect to the merits of potential estimators, we must think of the structural equations as an interconnected system. While we typically specify such equations from the perspective of determining their conditional means, the structural equations along with our independence assumptions determine the variances of the variables in the system. Above, this necessitates increasing the strength of the edge $\boldsymbol{U}\rightarrow \boldsymbol{A}$ while holding all other edges constant, which requires us to re-normalize the variances to maintain the strength of the edge $\boldsymbol{BAV} \rightarrow \boldsymbol{A}$.\\
\indent In appendix section \ref{sbsec: simulation invariant bav est}, we consider the properties of a simulation experiment aiming to vary the strength of the edge $\boldsymbol{BAV} \to \boldsymbol{A}$. We show that in the case that we fail to fix the variance of the treatment that the bias of the conditional estimator $\bav$ is invariant to $\gamma_{bav}, \forall \gamma_{bav} \in (-\infty, \infty)$,  but that the naive estimator is strictly increasing in $\gamma_{bav}$. It is clear from the theory we developed in section \ref{sc:variance accounted for} that if we increase the edge from $\boldsymbol{BAV} \to \boldsymbol{A}$ that amplification should strictly increase, but if we allow the variance in $\boldsymbol{A}$ to increase as the parameter increases, the amplification effect is precisely cancelled out.\\
\indent In general terms, simulating linear systems of location-scale family random variables requires first fixing the variances of the variables in the DAG. The relevant quantity determining the strength of the various edges are ratios of variances and covariances of the upstream parent nodes to the variance of the child node in determining the edge's strength. Since the effects are relative, in a simulation context we can normalize the variances to 1 or set them to the expected/observed variances of the data in a particular context. For simplicity we will demonstrate the normalized approach. In Figure \ref{fig:DAG1}, this means that $\sigma_u^2 = \sigma_a^2 = \sigma_{bav}^2 = \sigma_y^2 = 1$.\\
\indent The second step is to be explicit about independence and conditional independence assumptions. Given the independence assumptions, we can specify the covariance matrix of each child variable $\boldsymbol{Y_{child}}$ in terms of the matrix of $k$ arbitrary parent variables which form the edges going into the child variable, $\boldsymbol{{Y_{child}}}$.
\begin{align*}
\boldsymbol{Y_{child}} &= \boldsymbol{X_{parent}}\boldsymbol{\beta_{parent}} + \boldsymbol{\epsilon_{child}}\\
\implies Var(Y_{child}) &= \boldsymbol{\beta_{parent}^T}\boldsymbol{Var(X_{parent})}\boldsymbol{\beta_{parent}} + Var(\boldsymbol{\epsilon_{child}})
\end{align*}
\small
\begin{align*}
\sigma_{y_{child}^2} = 1 = \begin{bmatrix} \beta_{1} & \beta_{2} &\dots \beta_{k} \end{bmatrix}\begin{bmatrix} 1 & \sigma_{1,2} &\dots & \sigma_{1,k}\\
 \sigma_{1,2} & 1 & \dots & \sigma_{2,k} \\
\vdots & \sigma_{j,2} & \ddots & \vdots \\
\sigma_{k,1} & \sigma_{k,2} & \dots & 1\end{bmatrix} \begin{bmatrix} \beta_{1}\\
\beta_{2}\\
\vdots \\
\beta_{k}
\end{bmatrix} + \sigma_{\epsilon_{c}}^2
\end{align*}
\normalsize

The diagonal of all the parent covariance matrices is 1 since we have normalized all variables pictured in the DAG. The covariances themselves will be determined by the independence assumptions, the edges connecting the child nodes, and their structural equations. Essentially we are choosing the proportion of the child variation that the variances and the covariances of the parent variances explain. The error terms, $\epsilon$'s are the only non-normalized variances, and they absorb the shocks when we increase and decrease the strength of the edges of the non-error variables. This maintains the strength of all other relations visualized on the DAG.\\
\indent Since all variance terms must be non-zero (or equivalently that $\boldsymbol{\beta_{parent}^TVar(X_{parent})\beta_{parent}} \leq 1 = \sigma_{child}^2$), the variance equations define bounds on the simulation parameter space. In the above example, conditional on holding the strength of the edges $\boldsymbol{U} \rightarrow \boldsymbol{Y}, \boldsymbol{BAV} \rightarrow \boldsymbol{A}, \boldsymbol{BAV} \rightarrow \boldsymbol{Y}, \boldsymbol{A} \rightarrow \boldsymbol{Y}$, $\gamma_u \in (-0.893,0.893)$ defines the feasible range. That is, the edge $U\rightarrow A$ can explain up to 79.75\% of the variation in $\boldsymbol{A}$ ($\frac{\gamma_u^2\sigma_u^2}{\sigma_a^2} = \frac{\gamma_u^2}{1}$) since the edge $\boldsymbol{BAV}\rightarrow \boldsymbol{A}$ explains 20.25\% of the variation already. In general, the extent to which an edge can explain variation in the child node is constrained by the other child nodes and the covariance structure between those variables. A parameter however, such as $\gamma_u$ may be constrained by more than one set of inequalities. In this particular case $\gamma_u$ has to satisfy the following inequalities:

\begin{align*}
|\gamma_u| &\leq (1 - \gamma_{bav}^2)^\frac{1}{2}\\
\gamma_u & \leq \frac{1 - \beta_a^2 - \beta_u^2 - \beta_{bav}^2 - 2\beta_a\beta_{bav}}{2\beta_a\beta_u}
\end{align*}

\noindent where conditional on the strength of the particular edges ($\boldsymbol{U}\rightarrow \boldsymbol{Y}$, $\boldsymbol{BAV} \rightarrow \boldsymbol{A}$, $\boldsymbol{BAV} \rightarrow \boldsymbol{Y}$, $\boldsymbol{A}\rightarrow \boldsymbol{Y}$) in the above simulation, only the first inequality was binding.\\
\indent The nuance here is that the extent to which we can simulate unmeasured confounding depends upon not only how much amplifying we have simulated, but also on the true effect of the treatment on the outcome $\boldsymbol{A} \rightarrow \boldsymbol{Y}$. Since this is an interdependent system of equations, all of the parameters are competing for shares of fixed variances. If the treatment, independent of $\boldsymbol{U}$ and $\boldsymbol{BAV}$, explains the large majority of the outcome variance (i.e, the edge $\boldsymbol{A}\rightarrow \boldsymbol{Y}$), it means the weight of the edge $\boldsymbol{U} \rightarrow \boldsymbol{Y}$ must be relatively small, opposite signed, or the structural equations contain an effect modifier. This in turn constrains $\gamma_u$.\\
\indent Consider again the above simulation experiment where we are interested in varying the strength of $\boldsymbol{U}\rightarrow \boldsymbol{A}$ conditional on all other pathways. Suppose that the pathway $\boldsymbol{A} \rightarrow \boldsymbol{Y}$ explains $64\%$ of the variance in $\boldsymbol{Y}$, i.e that $\beta_a = 0.8$. Now both constraints on $\gamma_u$ are binding and the simulation parameter space is $\gamma_u \in (-0.893, 0.3916)$.\\
\indent In summary, when simulating linear location-scale family systems of equations we start by identifying the DAG and the independence assumptions between variables. Second, our simulation experiment should attempt to answer a causal question about how a proposed estimator behaves in response to an intervention on the weights of causal DAG. Just like experimental design, properly estimating the relevant counterfactural requires that the  difference in distributions between our intervention(s) and the control is the effect of the intervention(s) themselves. As demonstrated in this section, simluating linear systems of equations requires varying one of the edges of the DAG holding all else constant, and matching the means and variances of the simulated variables with that of the target observational study we are trying to mimic. This allows us to generate simulations whose distributions are proper counterfactuals. Third, conditional on the other edges, the covariance matrices impose bounds for the parameter space that we can simulate and thus the extent to which we can vary the edge of interest. For a specific realization of the experiment and accompanying valid parameters, the variables are constructed in the downstream direction, that is from parent nodes to child.\\
\indent In the example of simulating the proper intervention in Figure \ref{extended 1 squiggle}, we first simulate $\boldsymbol{U}$ and $\boldsymbol{BAV}$ independently with variance 1 respectively. Given $\gamma_u$ and $\gamma_{bav}$, the variance of the error term $\boldsymbol{\epsilon_2}$ from equation \ref{A truth} is implied and can be simulated. Having $\boldsymbol{U}$, $\boldsymbol{BAV}$ and $\boldsymbol{\epsilon_2}$ allows us to simulate the treatment $\boldsymbol{A}$. Conditional on the already simulated variables, their associated parameters, and $\beta_a$, $\beta_{bav}$, and $\beta_u$, the variance of the error term $\boldsymbol{\epsilon_1}$ is implied and can be simulated. Finally, since all of the child variables for the outcome have been simulated, we can simulate the outcome. To be clear, we can fix proportions of variance explained by each edge in any order we'd like as long as we respect the underlying constraints. However, given an admissible set of weights of the edges we must proceed from parent to child nodes to conduct the simulation.\\
\indent While this method requires us to calculate inequalities and make explicit the implied variance formulas for our variables, the benefits are that we can view our simulation as a well-defined causal experiment matching the constraints of our target study and we get sets of parameter bounds. When we do not keep the variance fixed, there are no defined bounds beyond heuristics, and more importantly, we are no longer matching the data to our target observational study. In many small systems, such as the one in Figure \ref{ext: 0 squiggle}, it is often computationally inexpensive to simulate a discretized approximation to all possible parameter configurations. In extremely large systems we can use domain knowledge to make refinements on these bounds and simulate a reasonable subset of the parameter space. This method allows us to make refinements over edges with strong priors while simulating the entirety of edges with greater uncertainty.

\section{Simulating Bias Amplification from a Real Data Set}\label{sc:real data}

Here we conduct a data simulation for an observational study. We want to consider a medical example with realistic amounts of variance in the treatment and the outcome. Further, we specifically consider the case of a binary treatment which is common in medical applications, biostatistics, and epidemiology. The difficulty, in general, when simulating with real data is that you do not know the true underlying parameter values. In this section, we start with a randomized controlled trial (RCT) and modify it appropriately, so that we can take the intention to treat (ITT) estimate as the true underlying effect for the foundation of our simulations.\\ 
\indent In our simulation experiment, we keep the treatment data unchanged (thus fixing their variance), and then simulate unmeasured confounding ($\boldsymbol{U}$) and bias amplifiers ($\boldsymbol{BAV}$) in order to modify selected covariates ($\boldsymbol{X}$) and the outcome ($\boldsymbol{Y}$) to produce a synthetic observational experiment. In order to precisely control the relationships between the simulated variables and the real variables we treat the binary treatment, $\boldsymbol{A}$, as though it comes from a latent probit model. 

\begin{align*}
\boldsymbol{A} &= 1(\boldsymbol{A^\star} >\boldsymbol{0}) \\
&= 1(\alpha_a + \boldsymbol{U}\gamma_u + \boldsymbol{\tilde{X}}\gamma_{\Tilde{x}} + \boldsymbol{\epsilon_2} > \boldsymbol{0})
\end{align*}

\noindent where $\boldsymbol{\tilde{X}} = \frac{\boldsymbol{X}}{\sigma^\prime} + \boldsymbol{BAV}$, and $\sigma^\prime$ is a scaling variable such that $\boldsymbol{X}$ and $\tilde{\boldsymbol{X}}$ have the same population variance. All of the latent variables ($\boldsymbol{U}$, $\boldsymbol{BAV_{n\times k}}$, $\boldsymbol{\epsilon_2}$, and hence $\boldsymbol{A^\star} = \alpha_a + \boldsymbol{U}\gamma_u + \boldsymbol{BAV}\gamma_{bav} + \boldsymbol{\epsilon_2} > \boldsymbol{0}$) are set to come from normal distributions. The details of the how the simulation is performed are in appendix section \eqref{real sim details}.\\
\indent For this paper, we use data from \cite{R2KJHK_2019}, a published RCT with 294 participants and relatively balanced distribution of covariates. While the reseachers examined many outcomes we will focus on the effects of an e-Health intervention in infants on child eating behaviours. The researchers gave the parents in the treatment group access to a "monthly age-appropriate video addressing infant feeding topics together with corresponding cooking films/recipes", and the outcome was eating habits of the child at a later point in time. In the observational study that we want to create,  (target observational study) we want to estimate the effect of the treatment on emotional overeating as measured by the Child Eating Behavior Questionnaire (CEBQ).

\subsection{Unbiased ITT Model}\label{sbsc: unbiased}

Our foundation is the unbiased ITT effect from the RCT data regressing the treatment on the outcome ($Y \sim A$) shown in the first column of Table \ref{regression table}.

\begin{center}
\begin{tabular}{@{}cccc@{}}
\toprule
Model                                 & ITT                          & ITT Cond.                    & $\boldsymbol{A}$     \\ \midrule
\multicolumn{1}{c|}{$\boldsymbol{A}$} & \multicolumn{1}{c|}{0.122}   & \multicolumn{1}{c|}{0.137}   & -                    \\
\multicolumn{1}{c|}{}                 & \multicolumn{1}{c|}{(0.052)} & \multicolumn{1}{c|}{(0.053)} &                    \\ \cmidrule(l){2-4} 
\multicolumn{1}{c|}{$\boldsymbol{CFNS}$}          & \multicolumn{1}{c|}{-}       & \multicolumn{1}{c|}{-0.007}  & -005                 \\
\multicolumn{1}{c|}{}                 & \multicolumn{1}{c|}{}        & \multicolumn{1}{c|}{(0.005)} & (0.007)              \\ \cmidrule(l){2-4} 
\multicolumn{1}{c|}{$\boldsymbol{CFQ}$}           & \multicolumn{1}{c|}{-}       & \multicolumn{1}{c|}{0.058}   & 0.036                \\
\multicolumn{1}{c|}{}                 & \multicolumn{1}{c|}{}        & \multicolumn{1}{c|}{(0.036)} & (0.040)              \\ \cmidrule(l){2-4} 
\multicolumn{1}{c|}{$\boldsymbol{Age_{mother}}$}           & \multicolumn{1}{c|}{-}       & \multicolumn{1}{c|}{0.008}   & 0.009                \\
\multicolumn{1}{c|}{}                 & \multicolumn{1}{c|}{}        & \multicolumn{1}{c|}{(0.006)} & (0.007)              \\ \midrule
$\mathcal{R}^2$                       & 0.018                        & 0.045                        & $\boldsymbol{0.009}$ \\ \bottomrule
\end{tabular}
\captionof{table}{Clinical Data Regression table. Each column represents a different regression. Column 1 is the unbiased Intention to Treat (ITT) model, regressing the outcome on the treatment. Column 2 represents the unbiased conditional ITT model with 3 covariates conditioned on. The third regression is the regression of the treatment on the 3 regressors. The row names are the independent variables, where $\boldsymbol{A}$ is the treatment, $CFNS$ is the Child Food Neophobia Score, $\boldsymbol{CFQ}$ is the Child Feeding Questionaire pressure subscale and  $\boldsymbol{Age_{mother}}$ is the age of the infant's mother. In brackets the relevant standard errors are displayed.}
\label{regression table}
\end{center}

\begin{figure}[H]
\centering
\begin{subfigure}{.44\textwidth}
    \centering
    \includegraphics[height = .14\textheight]{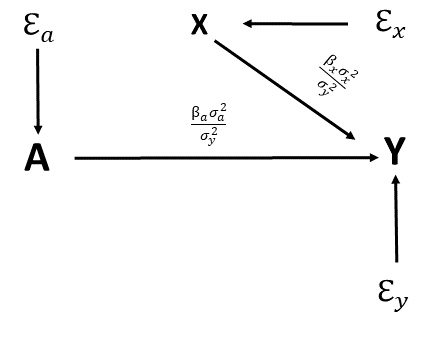}
    \caption{}
    \label{fig:DAG exp 1}
    \end{subfigure}
    \begin{subfigure}{.44\textwidth}
      \centering
    \includegraphics[height = .14\textheight]{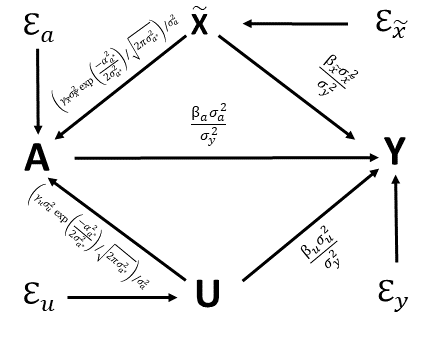}
    \caption{}
    \label{fig:DAG exper 2}
    \end{subfigure}
     \begin{subfigure}{.44\textwidth}
      \centering
    \includegraphics[height = .14\textheight]{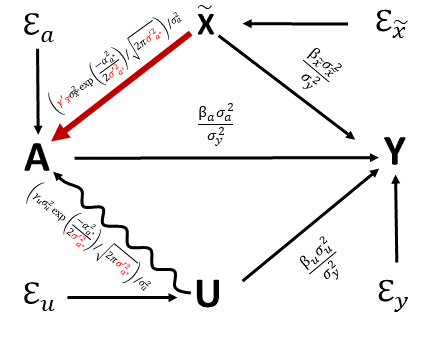}
    \caption{}
    \label{fig:DAG exper naive intervention}
    \end{subfigure}
     \begin{subfigure}{.44\textwidth}
      \centering
    \includegraphics[height = .14\textheight]{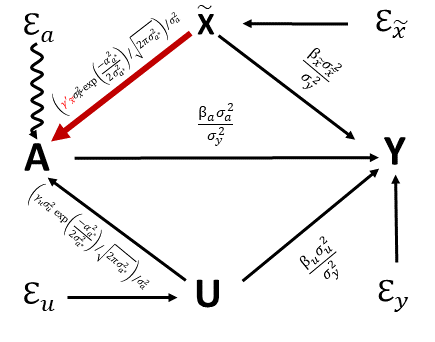}
    \caption{}
    \label{fig:DAG exper correct}
    \end{subfigure}
    \caption{Causal extended diagrams. (a) is the extended DAG representing the original experiment data. (b) represents the extended DAG with the modified data (see appendix \ref{real sim details} for details). (c) represents intervening on the causal DAG in (b) by changing the strength of the edge $\boldsymbol{\tilde{X}} \to \boldsymbol{A}$ without holding the latent treatment variance $\boldsymbol{A^\star}$ constant. The edges which have been modified inadvertently are shown as squiggly arrows. Parameters which have been changed are shown in red. (d) shows the intervention on the edge $\boldsymbol{\tilde{X}} \to \boldsymbol{A}$ while holding the latent variance constant.}
\end{figure}

In column 1 of table \ref{regression table}, we see that the ITT estimate is $0.12$. As this is an RCT, we do not expect baseline covariates [Child Food Neophobia Score ($\boldsymbol{CFNS}$), Child Feeding Questionaire ($\boldsymbol{CFQ}$) subscale pressure, and Age of mother ($\boldsymbol{Age_{mother}}$)] to be associated with exposure. We thus assume that the experimental data are generated from the causal DAG in Figure \ref{fig:DAG exp 1}, where $\boldsymbol{X}$ represents the matrix of all three covariates ($\boldsymbol{CFNS}$,$\boldsymbol{CFQ}$, and $\boldsymbol{Age_{mother}}$) after they have been individually standardized to have mean 0 and variance 1. To verify that these variables are not bias amplifiers, that is explain only a negligible proportion of the treatment variance, we also present the results of the regression of the treatment on the 3 covariates in column 3 in table \ref{regression table}. We can see that jointly and individually the three covariates explain very little of the variance in the treatment, $\mathcal{R^2} = 0.009$. This should be expected in a truly randomized experiment set-up since proper randomization breaks the causal association from the covariates to the treatment.

\indent Since these covariates do not cause $\boldsymbol{A}$ and we have assumed that the ITT estimator is unbiased, when we estimate $\boldsymbol{Y} \sim \boldsymbol{A} + \boldsymbol{CFNS_{score}}+ \boldsymbol{CFQ_{pressure}}+ \boldsymbol{Age_{mother}}$ the expectation and probability limit of $\hat{\beta_a}$ remains unchanged regardless of the strength of association between the covariates and the outcome. However, actual results may vary due to final sample variation. In our RCT data, the unadjusted model estimates a treatment effect of $0.122$ and the adjusted model estimates $0.137$. Since simulation experiments performed in section \ref{sbsec: real data sim main} all condition on covariates, we consider the covariate adjusted results from the RCT as the gold standard for determining bias due to unmeasured confounding in our simulated data.

\subsection{Biased Model Simulations}\label{sbsec: real data sim main}

 Our objective is to simulate data according to the DAG in Figure \ref{fig:DAG exper 2}. To produce the simulations, we took 10000 bootstrap replications of the original outcome, treatment and covariates. From each bootstrap sample of the treatment, $\boldsymbol{A_{bootstrap}}$, of size $n=294$ we simulated the latent variable $\boldsymbol{A^\star}$ using the procedure outlined in the appendix (section \ref{real sim details}). Next, conditional on the drawn latent samples of $\boldsymbol{A^\star}$ and the bootstrapped covariates, we drew samples for the unmeasured confounding, $\boldsymbol{U}$, and bias amplifying variable, $\boldsymbol{BAV}$. The modified random control variables, $\boldsymbol{\Tilde{X}} = \frac{\boldsymbol{X}}{\sigma^\prime} + \boldsymbol{BAV}$, were produced by adding the bias amplifying variables to a scaled version of the original control variables. Linear combinations of the unmeasured confounding and modified covariates were then added with reasonable values to the outcome such that the following DAG and equations hold (see simulation results).

\begin{align}
    \boldsymbol{\Tilde{Y}} &= \alpha_y + \boldsymbol{A}\beta_a + \boldsymbol{\Tilde{X}}\beta_{\tilde{x}} + \boldsymbol{U}\beta_u + \boldsymbol{\epsilon_1}\label{outcome mod rct}\\
    \boldsymbol{A^\star} &= \alpha_a +\boldsymbol{U}\gamma_u + \boldsymbol{\Tilde{X}}\gamma_{\Tilde{X}} + \boldsymbol{\epsilon_2}\label{latent mod rct}\\
    \boldsymbol{A} &= 1\{\boldsymbol{A^\star} >\boldsymbol{0}\}\label{treatment mod rct}
\end{align}

In section \eqref{sbsc: unbiased} we showed that the true treatment effect was $0.137$ conditional on the covariates $\boldsymbol{X}$. In the boostrap simulation pictured below, the unbiased model conditional on both the modified covariates, $\boldsymbol{\Tilde{X}}$, and the unmeasured confounding $\boldsymbol{U}$ is $0.136$ as expected. The naive model estimator had an average estimate of $0.234$ in the simulations and thus an absolute estimated bias of $0.097$, or a relative bias of 1.8 standard deviations ($\frac{0.234 - 0.137}{0.053}$) with respect to the unbiased estimate in section \eqref{sbsc: unbiased}.\\
\indent When we further condition on the modified covariates, the absolute bias ($E[|\hat{\beta}_a^{|\tilde{x}}- 0.137|]$) more than doubles to 0.225, and the relative bias increases to 4.3 standard deviations ($\frac{0.36 - 0.137}{0.053}$) with respect to the unbiased estimate in section \eqref{sbsc: unbiased}. 

\begin{figure}[H]
    \centering
    \includegraphics[scale = .4]{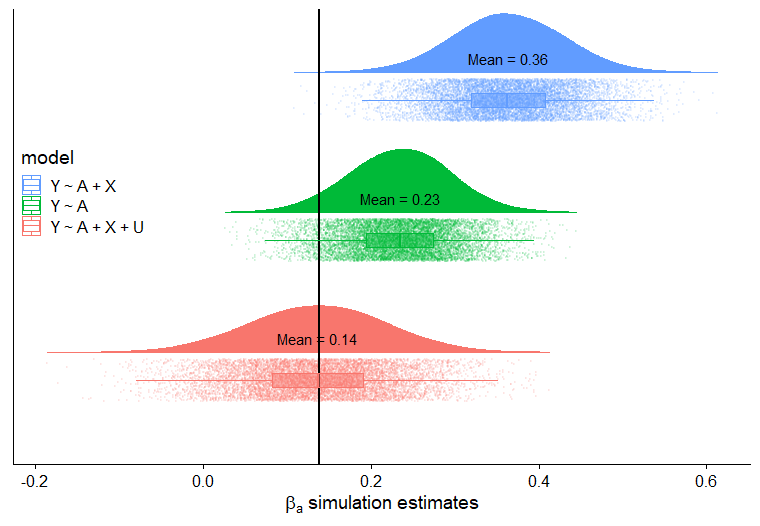}
    \caption{Here we compare three estimators for $\beta_a$ from the structural equations in equations \eqref{outcome mod rct}, \eqref{latent mod rct}, and \eqref{treatment mod rct}. In red, the results of the unbiased and infeasible estimator are shown, centered at the true value $\beta_a = 0.137$ (shown by the vertical black line). In green, the replications for the naive estimator ($\naive$) is shown and in blue the replications for the conditional estimator ($\hat{\beta}^{|\tilde{X}}$) is shown. Simulation details: Bootstrap replications = 10,000. $\beta_a = 0.137$, $\beta_u = 0.15$, $\boldsymbol{\beta_{\tilde{x}}} = (0.10, -0.15, -0.10)$, $COV(\boldsymbol{A},\boldsymbol{U})$ = 0.25,  $COV(\boldsymbol{A},\boldsymbol{\tilde{X}}) = (0.22, 0.15, 0.13)$}
    \label{rainforest_experiment}
\end{figure}

The simulations confirm that bias amplification can be significant even when constrained to problems of realistic variance. Further, we see that bias amplification is potentially a problem for binary outcomes. This underscores the theoretical points made in sections \eqref{sc:variance accounted for} and \eqref{sec:identify_amp} where we showed that the phenomenon behind bias amplification does not require specific distributional assumptions of the variables in the model.\\
\indent More importantly, by combining the methodology outline in the appendix (See \ref{real sim details}) to simulate measured confounding using real data and the principles for simulating systems of equations in section \ref{sc: new simulation}, we can produce realistic and complete simulations of parameter spaces which match the underlying characteristics of the data. Investigators who choose covariates based on the assumption of no unmeasured confounding can now evaluate the amount of bias amplification that would occur if this assumption does not hold.\\
\indent Finally, in the appendix (section \ref{sbsec: real sim estimator}) we consider an example of a causal simulation experiment with a binary treatment variable under the DAG in Figure \ref{fig:DAG exper 2} and structural equations \eqref{outcome mod rct}, \eqref{latent mod rct}, and \eqref{treatment mod rct}. The experiment involves modifying the strength of the edge $\boldsymbol{\tilde{X_1}} \rightarrow \boldsymbol{A}$ and evaluating the impact on the naive and conditional estimators.  With binary treatment ($\boldsymbol{A}$), we show that if we fail to hold the variance of the latent treatment ($\boldsymbol{A^\star}$) constant and increase $\gamma_{\tilde{x_1}}$, then it is possible to decrease the amount of observed treatment variance ($\sigma_{a}^2$) explained by $\boldsymbol{\tilde{X_1}}$. Further, the increased treatment variance also decreases the strength of the edge $\boldsymbol{U} \rightarrow \boldsymbol{A}$.   As a result of performing the causal simulation experiment improperly, it appears as though that varying the strength of the potential amplifiers has a negligible or negative impact on the resulting bias amplification. The improper and proper approaches to intervention are shown in Figure \ref{fig:DAG exper naive intervention} and Figure \ref{fig:DAG exper correct} respectively and the results from these simulations are visualized in the appendix in Figure \ref{fig:raincloud facet} in Appendix section \ref{sbsec: real sim estimator}. This of course leads to improper inferences regarding the relative merits of the naive and conditional estimators as well.  This highlights once again the importance of comparing simulations with comparable properties and ensuring that when we intervene on the edges of our causal diagram that we are not inadvertently varying the edges we mean to keep fixed. Just as in the experimental context, our simulation results become muddled or meaningless if we are not evaluating well-articulated counterfactuals.

\section{Discussion}

Causal model selection techniques have largely been developed under the assumption that a sufficient set of variables is available to create ignorability. When a sufficient set is not available or when a causal variable selection technique does not correctly identify the sufficient set, we are at risk of bias amplification. In the first simulation in section \ref{sec:intro}, we showed that even under mild perturbations of the usual assumptions, conditioning on a set of jointly strong proxy variables for $\boldsymbol{A}$ in OLS led to a very biased estimator (0.73 standard deviations on average). Further, most current causal variable selection techniques are likely to include this set of variables since they are significant predictors of the outcome and the treatment as well as variables which cause large changes in estimates when included sequentially.\\
\indent Under threat of bias amplification, treatment-oriented selection techniques for regression analyses using continuous exposure regimes should be used cautiously unless one has strong priors that a sufficient set is available and likely to be identified. We showed in section \ref{sc:variance accounted for} that it is precisely the amount of variance in the treatment explained by the observables in our model which is responsible for bias amplification. Similarly, we can see that a significant change in estimate is not sufficient to suggest that overall bias is decreasing since this could be the result of further bias amplification.\\
\indent These results call for new techniques to be developed for observational studies which can accommodate unmeasured confounding to help researchers choose reasonable and least-biased methods. We suggest to first identify the most plausible causal DAG. From the DAG and basic structural equation assumptions, an expression for asymptotic bias can often be derived. Further, we suggest to estimate the always-identifiable amplification term in observational settings and to assess the risk of bias amplification. With a measure for amplification and a limiting bias expression, a sensitivity analyses can be performed. One reasonable sensitivity analysis approach would be to estimate the amount of unmeasured confounding required in the spirit of E-values \citep{VanderWeelePeng_Sensitivity2017} to determine the strength of confounding associations required to "explain away the treatment effect" \citep{VanderWeelePeng_Sensitivity2017} and to make principled inferences from the data. This would require, as we have shown, properly simulating the unmeasured confounded as to respect the properties of the original data and such that the other competiting effects, i.e edges of the DAG, are not inadvertently altered. In such a set-up, large effects and relatively small amplifying terms lend credibility to results as being robust to unmeasured confounding, particularly in cases when suitable priors can be placed on the variables along the unmeasured confounding pathway. Alternatively, one could follow the approach of \cite{hill_sensitivity} and use the underlying structural equations and the data to generate candidate values of the unmeasured confounding. As we showed in section \ref{sc: new simulation} it is important that any such simulation method take into account the asymmetry of bias amplification with respect to the weight of the edge $\boldsymbol{U} \to \boldsymbol{A}$ and $\boldsymbol{U} \to \boldsymbol{Y}$.\\
\indent Ultimately, simulation experiments must aim to produce data from which we can draw causal conclusions to questions about estimators or functions. This means having well-defined interventions on the edges of the causal graphs and holding the other edges constant. In linear systems of equations, this requires keeping the moments of the variables, in particular variance, fixed when modifying the  weight of the DAG's edges. If we allow the treatment variance to vary incidentally as we increase confounding effects, the intervention arm of our simulations will no longer match the target observational study in the control arm. As a further consequence, the additional variance in the exposure may absorb much of the amplifying effect. This leads to systematic underestimation of bias amplification and may be an explanation for why the threat of bias amplification has not been appreciated as a concern for applied researchers\cite{meyers2011}. Fixing the variance of the variables has the additional benefit of defining the feasible parameter space. By constraining the underlying parameters by the implied variance equations (e.g equation (\ref{eq:variance determination})), it is computationally and conceptually easier to simulate the entire range of plausible treatment effects and biases. This leads to more representative simulations and more principled inferences.
\begin{dci}
The author(s) declared no potential conflicts of interest with respect to the research, authorship, and/or publication of this
article.
\end{dci}
\begin{funding}
The authors received funding from the Canadian Institute of Health Research (CIHR) through the Collaborative Health Research Projects (NSERC partnered) for the research and publication of this article. Grant number: CPG-140204.
\end{funding}


\nocite{*}
\bibliography{main}
\bibliographystyle{vancouver}


\appendix
\section{Appendix}
\label{sc:Appendix}

\subsection{Matrix Notation and FWL Theorem}\label{sbsec: matrix and fwl}
\label{sbsec:appendix-matrix}

Throughout this paper, we make use of matrix notation to concisely represent estimates and as a way of considering the geometry of the least squares. Here is a quick guide for understanding the notation in this paper.\\

Let $\boldsymbol{A}$ be the $n \times p$ matrix of treatment variables. For illustrative purposes consider that $A$ is a single binary $n\times 1$ vector.

\begin{align*}
\boldsymbol{A_{n\times1}} = \begin{bmatrix}
1 \\ 1\\ 0 \\1\\ \vdots \\0
\end{bmatrix}
\end{align*}

There are n rows of data, each with a $1$ or $0$ representing the observation being treated or not.\\

Another piece of notation that is used is annihilator and orthogonal projection matrices. Let $\boldsymbol{P_X}$ be the orthogonal projection matrix of $\boldsymbol{X}$, an $n\times k$ matrix, and $\boldsymbol{M_X}$ the annihilator or residual-making matrix of $\boldsymbol{X}$, 

\begin{align*}
\boldsymbol{P_X} &= \boldsymbol{X(X^TX)^{-1}X^T};\\
\boldsymbol{M_X} &= \boldsymbol{I - P_X} = \boldsymbol{I - X(X^TX)^{-1}X^T}.\\
\end{align*}

A projection matrix maps each point to the nearest point in the subspace spanned by the columns in $\boldsymbol{X}$, $S(\boldsymbol{X})$.The annihilator matrix maps each point to the orthogonal complement of $S(\boldsymbol{X})$, $S^\perp (\boldsymbol{X})$. The predicted outcome in ordinary least squares is $\boldsymbol{\hat{Y}} = \boldsymbol{X\hat{\beta}} = \boldsymbol{X(X^TX)^{-1}X^TY} = \boldsymbol{P_XY}$, which we can think of geometrically as "dropping a perpendicular" \citep{davidson2004econometric} from the outcome vector into the subspace spanned by the covariates in the regression. The orthogonal complement to the space spanned by the regressors, $S^\perp(\boldsymbol{X})$ is where the fitted residual vector, $\boldsymbol{\hat{\epsilon}}$, lives. We can see that the residual vector $\boldsymbol{\hat{\epsilon}} = \boldsymbol{Y} - \boldsymbol{\hat{Y}} = \boldsymbol{Y} - \boldsymbol{X\hat{\beta}} = \boldsymbol{Y} - \boldsymbol{P_XY} = \boldsymbol{(I - P_X)Y} = \boldsymbol{M_XY}$, is just the projection of Y into the subspace orthogonal to $S(\boldsymbol{X})$.\\

By definition, we can always then decompose $\boldsymbol{Y}$ uniquely into its projection onto $S(\boldsymbol{X})$ and $S^\perp(\boldsymbol{X})$, 

\begin{align*}
\boldsymbol{Y} = \boldsymbol{P_XY} + \boldsymbol{M_XY}. 
\end{align*}

Orthogonal projection matrices have two important properties, they are symmetric and idempotent. This means that $\boldsymbol{P_X} = \boldsymbol{P_{X}^T}$ and $\boldsymbol{P_XP_X} = \boldsymbol{P_X}$ and that these same two properties are equally enjoyed by $\boldsymbol{M_X}$. Further, any matrix in the subspace spanned by $\boldsymbol{X}$ is \textit{annhilated} when operated on by $\boldsymbol{M_X}$, since it is by definition orthogonal to $S(\boldsymbol{X})$.\\

We also appeal to the Frisch-Waugh-Lovell (FWL) theorem to construct the matrix notation regression estimates as well as for visualizing the 2 dimensional plot of a single regression in the context of a multivariable regression. Suppose we construct an arbitrary partition of $\boldsymbol{X_{n\times k}} = [\boldsymbol{X_{1_{n\times k_1}},X_{2_{n\times k_2}}}]$, where $k_1 + k_2 = k$. The FWL theorem states that the following two regressions,\eqref{eq: FWL 1} and \eqref{eq: FWL 2}  produce numerically equivalent estimates of the vector $\boldsymbol{\hat{\beta_2}}$ as well as numerically equivalent residuals.

\begin{align}
    \boldsymbol{Y} &= \boldsymbol{X_1\beta_1} + \boldsymbol{X_2\beta_2} + \boldsymbol{\epsilon}\label{eq: FWL 1}\\
    \boldsymbol{M_{X_1}Y} &= \boldsymbol{M_{X_1}X_1\beta_1} + \boldsymbol{M_{X_1}X_2\beta_2} + \boldsymbol{M_{X_1}\epsilon}\\
    &= \boldsymbol{M_{X_1}X_2\beta_2} + \boldsymbol{M_{X_1}\epsilon}\label{eq: FWL 2}
\end{align}

What this says in words is that it is numerically equivalent to regress $\boldsymbol{Y}$ on the columns of $\boldsymbol{X_1}$ and $\boldsymbol{X_2}$ simultaneously as it is to first regress both $\boldsymbol{Y}$ and $\boldsymbol{X_2}$ on the columns of $\boldsymbol{X_1}$ separately, then save the respective residuals, $\boldsymbol{M_{X_1}Y}$ and $\boldsymbol{M_{X_1}X_2}$, and regress the former on the later. By simply pre-multiplying both sides of \eqref{eq: FWL 2} by $\boldsymbol{X_2^T}$ and rearranging, we get the general matrix notation formulation for the vector of ceofficient estimates $\boldsymbol{\hat{\beta_2}} = \boldsymbol{(X_2M_{X_1}X_2)^{-1}X_2M_{X_1}Y}$. Using the idempotency and symmetry properties we can rewrite the coefficient estimate:

\begin{align}
   \boldsymbol{\hat{\beta_2}} &= \boldsymbol{(X_2^TM_{X_1}X_2)^{-1}X_2^TM_{X_1}Y}\\
   &= \boldsymbol{\frac{(M_{X_1}X_2)^T(M_{X_1}Y)}{(M_{X_1}X_2)^T(M_{X_1}X_2)}}\\
   &=\boldsymbol{\frac{\hat{\upsilon}_{x_2}\hat{\upsilon}_{y}}{\hat{\upsilon}_{x_2}^2}}
\end{align}
\noindent where $\boldsymbol{\hat{\upsilon}_{x_2}}$ and $\boldsymbol{\hat{\upsilon}_y}$ are the residuals from the regression of $\boldsymbol{X_2}$ and $\boldsymbol{Y}$ on $\boldsymbol{X_1}$ respectively.\\

A special case of the above result is when we have a $n\times k_1$ matrix of treatment variables, $\boldsymbol{A}$ and a $n\times k_2$ matrix of controlling variables. For example, we are trying to estimate the causal effect of the matrix $\boldsymbol{A}$ on the outcome $\boldsymbol{Y}$ using a selection on observables strategy by conditioning on $\boldsymbol{Z}$. The FWL theorem tells us that the estimates of the causal effect, $\boldsymbol{\hat{\beta}_a}$ can be obtained by the two following regression equations:

\begin{align*}
\boldsymbol{Y} &= \boldsymbol{A\beta_a} + \boldsymbol{Z\beta_z} + \boldsymbol{\upsilon_1}\\
\boldsymbol{M_ZY} &= \boldsymbol{M_ZY} + \boldsymbol{\upsilon_1} \label{FWL simple}
\end{align*}

The second regression is a simple linear regression, with only one dependent variable $\boldsymbol{M_ZY}$ and a single regressor, $\boldsymbol{M_ZA}$. The error term remains unchanged by the projection into $S^\perp(\boldsymbol{Z})$ since it can be represented as $\boldsymbol{M_{A,Z}Y}$ which is already contained in the subspace $S^\perp(\boldsymbol{Z})$. Another way we can write the estimate, $\boldsymbol{\hat{\beta_a}}$ is to apply the well known $\boldsymbol{\hat{\beta}} = \boldsymbol{(X^TX)^{-1}X^TY}$  to regression equation \eqref{FWL simple}.

\begin{align*}
\boldsymbol{\hat{\beta}_a} &= \boldsymbol{\frac{(M_ZA)^T(M_zY)}{(M_ZA)^TM_ZA}}\\
&= \boldsymbol{\frac{\frac{1}{n}(M_ZA)^T(M_zY)}{\frac{1}{n}(M_ZA)^TM_ZA}}
\end{align*}

The numerator $\frac{1}{n}\boldsymbol{(M_ZA)^T(M_zY)}$ can be seen as the dot product of the residuals from the regression of the treatment on the control variables, $\boldsymbol{A} \sim \boldsymbol{Z}$, and the residuals from the regression of the outcome on the control variables, $\boldsymbol{Y} \sim \boldsymbol{Z}$, scaled by $\frac{1}{n}$. If a column of ones is included in the matrix $\boldsymbol{Z}$, both sets of residuals will be centered. We can then think of the dot product in the numerator as an estimator for the covariance of the two residuals, $COV(\boldsymbol{\epsilon_a},\boldsymbol{\epsilon_y})$. In general terms we will have bias due to unmeasured confounding if the covariance is a function of $\boldsymbol{U}$. The denominator can be seen as numerically equal to the sum of squared residuals.\\

An important special case of the annhilator matrix is $\boldsymbol{M_{\iota}}$, where $\boldsymbol{\one}$ is a $n\times 1$ vector of ones. This is sometimes called the centering matrix because it de-mean's the matrix it operates on, since  $\boldsymbol{\miota X} = \boldsymbol{X} - \boldsymbol{1}(\boldsymbol{1}^T\boldsymbol{1})^{-1}\boldsymbol{1}^T\boldsymbol{X} = \boldsymbol{X} - \frac{1}{n}\sum_{i=1}^nx_i = \boldsymbol{X}-\boldsymbol{\bar{X}}$.\\

Using the symmetry and idempotency properties combined with the convergence in probability properties discussed in the next subsection this implies:
\begin{align}
    \frac{1}{n}\boldsymbol{X_1\miota X_2} &= \frac{1}{n}(\boldsymbol{X_1} - \boldsymbol{\bar{X_1}})(\boldsymbol{X_2} - \boldsymbol{\bar{X_2}})\\
    &= \frac{1}{n}\sum_{i=1}^n(x_{1_i} -\boldsymbol{\bar{X_1}})(x_{2_i} -\boldsymbol{\bar{X_2}})\\
    &\overset{p}\to E[(\boldsymbol{X_1} - E[\boldsymbol{X_1}])(\boldsymbol{X_2} - E[\boldsymbol{X_2}])\\
    &= COV(\boldsymbol{X_1},\boldsymbol{X_2})
\end{align}
When $\boldsymbol{X_1} = \boldsymbol{X_2}$, the last line becomes $\boldsymbol{Var(X_1)}$.


For a more complete and technical treatment of projection and annihilation matrices as it pertains to OLS see Econometric Theory and Methods by Davidson and MacKinnon \cite{davidson2004econometric}.

\subsection{Convergence in Probability}\label{sbsec:convergence in probability}
Throughout the paper we use the notation $\plim_{n\to \infty}$ to mean the limit in probability. Specifically, if $\plim_{n\to \infty} \boldsymbol{Y_n} = \boldsymbol{Y}$ then:
\begin{align*}
lim_{n\to\infty} P(|\boldsymbol{Y_n} - \boldsymbol{Y}|>\epsilon) = 0, \forall \epsilon > 0
\end{align*}

Alternatively we can write $\plim_{n\to \infty} \boldsymbol{Y_n} = \boldsymbol{Y}$ as $\boldsymbol{Y_n} \overset{p}\to \boldsymbol{Y}$ or simply $\plim$. Throughout this paper, all probability limits are as $n \to \infty$.\\

Below are a few important properties of Probability limits used throughout the paper. Suppose $\boldsymbol{X_n} \overset{p}\to \boldsymbol{X}$ and $\boldsymbol{Y_n} \overset{p} \to \boldsymbol{Y}$ then:

\begin{align}
\boldsymbol{X_n} + \boldsymbol{Y_n} \overset{p}\to \boldsymbol{X} + \boldsymbol{Y}\label{plim add}\\
\boldsymbol{X_n Y_n} \overset{p}\to \boldsymbol{X Y} \label{plim mult}\\
\frac{\boldsymbol{X_n}}{\boldsymbol{Y_n}} \overset{p}\to \frac{\boldsymbol{X}}{\boldsymbol{Y}}\label{plim div}
\end{align}
\noindent where the third line is just a special case of the second and holds whenever the denominator is well defined. These properties are well known and follow from the Continuous Mapping Theorem.\\

Another useful theorem we use in this paper is the Weak Law of Large Numbers (WLLN). Here we consider a set of standard assumptions. Suppose we take the sample average of random variables $\{\boldsymbol{X_1},\boldsymbol{X_2},...,\boldsymbol{X_n}\}$, such that $\boldsymbol{X_i}$'s are independent and identically distributed (iid) and $E[\boldsymbol{X_i}] = m < \infty$, i.e the expectation is finite then:

\begin{align}
 \frac{1}{n}\sum_{i=1}^n\boldsymbol{X_i} \overset{p}\to E[\boldsymbol{X_i}] = m.
\end{align}

In the paper, whenever specified we assume that the error terms are coming from a normal distribution. In light of the WLLN, we can see that normal error terms are not required for the results to hold, that in fact we just need the error terms to come from an identical and independent distribution. The above results can be weakened further such that we can replace the iid condition with pairwise independence (See \cite{chung2001course} for details).\\

Further, some probability limit results do not always have a closed form expression, for example \ref{plim:cond nonlinear}. Sometimes we express the resulting limit as a function of random variables. These random variables tend to their respective probability limits, provided they exist.

\subsection{Derivations Continued}\label{subsc:derivations}

Below are derivations, extensions, proofs, and alternate forms of the equations presented in the main text.

\subsubsection{Average Causal and Average Partial Effects}

Throughout this paper we consider linear models with continuous exposures and as such a natural causal estimand of interest is the Average Partial Effect (APE). Under the linearity assumptions, the Average Partial Effect coincides with the Average Causal Effect. Below we show the derivation of the APE under the various DAG and structural equation assumptions. Implicitly, we further assume standard regularity conditions such as existence and boundedness of the estimators in $\mathcal{L}_1$, so that the derivative operator can freely move inside the expectation integral.

\subsubsection{Average Partial Effects for equations \eqref{Y truth} and \eqref{A truth}}
\small
\begin{align*}
\text{APEs  } &= \frac{\partial E[\boldsymbol{Y}|\boldsymbol{A},\boldsymbol{U},\boldsymbol{BAV}]}{\partial \boldsymbol{A}}\\
 &=  \frac{\partial E[\alpha_y + \boldsymbol{A\beta_a} + \boldsymbol{U\beta_u} + \boldsymbol{BAV\beta_{bav}} + \boldsymbol{\epsilon_1}|\boldsymbol{A,U,BAV}]}{\partial \boldsymbol{A}}\\
 &= \beta_a
\end{align*}
\normalsize
When we allow for $\boldsymbol{BAV}$ to be a $n \times k$ vector and for $\boldsymbol{\beta_{BAV}}$ to be potentially a zero vector, we can see that the above derivation holds for all of the DAG's and structural equations which assume there is no interaction term.

\subsubsection{$\hat{\beta}_a^{naive}$ in equation \ref{beta naive 1}}

\begin{align}
\hat{\beta}_a^{naive} &= \boldsymbol{\frac{A^TM_zY}{A^TM_zA}}\\
&= \beta_a + \beta_u\boldsymbol{\frac{A^TM_{\boldsymbol{1}}U}{A^TM_{\boldsymbol{1}}A}} + \beta_{bav}\boldsymbol{\frac{A^TM_{\boldsymbol{1}}BAV}{A^TM_{\boldsymbol{1}}A}}+\boldsymbol{\frac{A^TM_{\boldsymbol{1}}\epsilon_1}{A^TM_{\boldsymbol{1}}A}}\label{beta naive 2}\\
&= \beta_a + \beta_u\frac{\frac{1}{n}\sum_{i=1}^{n}(a_i - \bar{a})(u_i - \bar{u})}{\frac{1}{n}\sum_{i=1}^{n}(a_i - \bar{a})^2} + \beta_{bav}\frac{\frac{1}{n}\sum_{i=1}^{n}(a_i - \bar{a})(iv_i - \bar{bav})}{\frac{1}{n}\sum_{i=1}^{n}(a_i - \bar{a})^2} +
 \frac{\frac{1}{n}\sum_{i=1}^{n}(a_i - \bar{a})(\epsilon_{1_i} - \bar{\epsilon_1})}{\frac{1}{n}\sum_{i=1}^{n}(a_i - \bar{a})^2}\label{beta naive 3}
\end{align}

\subsection{Derivation of Pearl (2011) result}\label{pearl derivation}

Here we will explicitly follow Pearl's derivation \citep{pearl2012class}, to show the advantages of considering the probability limit over strictly expectations. We will derive the expectation for $\bav$ from estimating \eqref{seq iv}, which is the Average Partial Effect conditional on the treatment, A, and the BAV variable.

\begin{align*}
E[\bav] &= \frac{\partial E[\boldsymbol{Y}|\boldsymbol{A,BAV}]}{\partial \boldsymbol{A}}\\
&= \frac{\partial E[\alpha_y + \boldsymbol{A}\beta_a + \boldsymbol{U}\beta_u +\boldsymbol{BAV}\beta_{bav} + \boldsymbol{\epsilon_1}|\boldsymbol{A,BAV}]}{\partial \boldsymbol{A}}\\
&= \beta_a + \beta_u\frac{\partial E[\boldsymbol{U}|\boldsymbol{A,BAV}]}{\partial \boldsymbol{A}}
\end{align*}

We must find the expectation of $\boldsymbol{U}$ conditional on $\boldsymbol{A}$ and $\boldsymbol{BAV}$. Pearl solves this challenge by supposing the true underlying relationship between $\boldsymbol{U}$, $\boldsymbol{A}$, and $\boldsymbol{BAV}$ is linear and writing this functional form as a linear regression equation:

\begin{align*}
\boldsymbol{U} = \alpha_u + \boldsymbol{A}\zeta_a + \boldsymbol{BAV}\zeta_{bav} + \boldsymbol{\epsilon_3}
\end{align*}
Using this equation in addition to the two structural equations for $\boldsymbol{Y}$ and $\boldsymbol{A}$ respectively, we can express the regression coefficients, $\zeta_a$ and $\zeta_{bav}$ in terms of the structural coefficients $\beta_a$, $\beta_u$, $\gamma_a$, $\gamma_u$ by equating expressions for the covariances under the two sets of structural equations.

\begin{align*}
COV(\boldsymbol{U},\boldsymbol{A}) &= E[\boldsymbol{AU}] - E[\boldsymbol{A}]E[\boldsymbol{U}]\\
&= E[(\alpha_a + \boldsymbol{U}\gamma_u + \boldsymbol{BAV}\gamma_{bav} + \boldsymbol{\epsilon_2})\boldsymbol{U}] - E[(\alpha_a + \boldsymbol{U}\gamma_u + \boldsymbol{BAV}\gamma_{bav} + \boldsymbol{\epsilon_2})]E[\boldsymbol{U}]\\
&= \gamma_u(E[\boldsymbol{U^2}] - E[\boldsymbol{U}]^2)\\
&= \gamma_u\sigma_u^2 \label{cov au}
\end{align*}
Equivalently
\begin{align*}
COV(\boldsymbol{U,A}) &= E[\boldsymbol{AU}] - E[\boldsymbol{A}]E[\boldsymbol{U}]\\
&= E[\boldsymbol{A}(\alpha_u + \boldsymbol{A}\zeta_a + \boldsymbol{BAV}\zeta_{bav})] - E[\boldsymbol{A}]E[(\alpha_u + \boldsymbol{A}\zeta_a + \boldsymbol{BAV}\zeta_{bav})]\\
&= \zeta_a(E[\boldsymbol{A^2}]-E[\boldsymbol{A}]^2) + \zeta_{bav}(E[\boldsymbol{A BAV}]
-E[\boldsymbol{A}]E[\boldsymbol{BAV}])\\
&= \zeta_a\sigma_a^2 + \zeta_{bav}(COV(\boldsymbol{A},\boldsymbol{BAV}))\\
&= \zeta_a\sigma_a^2 + \zeta_{bav}\gamma_{bav}\sigma_{bav}^2,
\end{align*}

where the last line follows analogously from our derivation of $COV(\boldsymbol{A},\boldsymbol{U})$. Putting these together we have that:

\begin{align}
\zeta_a &= \frac{\gamma_u\sigma_u^2 - \zeta_{bav}\gamma_{bav}\sigma_{bav}^2}{\sigma_a^2}. 
\end{align} 

Similarly, putting the two steps together for succinctness, 

\begin{align}
COV(\boldsymbol{U},\boldsymbol{BAV}) &= 0 \text{ (independence) }\\
&= E[\boldsymbol{U BAV}] - E[\boldsymbol{U}]E[\boldsymbol{BAV}]\\
&= E[(\alpha_u + \boldsymbol{A}\zeta_a + \boldsymbol{BAV}\zeta_{bav})\boldsymbol{BAV}] - E[(\alpha_u + \boldsymbol{A}\zeta_a + \boldsymbol{BAV}\zeta_{bav})]E[\boldsymbol{BAV}]\\
&= \zeta_aCOV(\boldsymbol{A},\boldsymbol{BAV}) + \zeta_{bav}Var(\boldsymbol{BAV})\\
&= \zeta_a\gamma_{bav}\sigma_{bav}^2 + \zeta_{bav}\sigma_{bav}.
\end{align}

Now we have two equations for the two new regression coefficients in terms of the structural equations. Combining 

\begin{align}
\zeta_a &= \frac{\gamma_u\sigma_u^2}{\sigma_a^2 - \gamma_{bav}^2\sigma_{bav}^2}\\
\zeta_{bav} &= \frac{-\gamma_u\gamma_{bav}\sigma_u^2}{\sigma_a^2 - \gamma_{bav}^2\sigma_{bav}^2}
\end{align}

Returning to the task of finding the partial effect of $\boldsymbol{A}$ on $E[\boldsymbol{Y}|\boldsymbol{A,BAV}]$ and thus $E[\bav]$.

\begin{align}
E[\bav] &= \frac{\partial E[\boldsymbol{Y}|\boldsymbol{A},\boldsymbol{BAV}]}{\partial \boldsymbol{A}}\\
&= \frac{\partial E[\alpha_y + \boldsymbol{A}\beta_a + \boldsymbol{U}\beta_u +\boldsymbol{BAV}\beta_{bav} + \boldsymbol{\epsilon_1}|\boldsymbol{A,BAV}]}{\partial \boldsymbol{A}}\nonumber\\
&= \beta_a + \frac{\partial \beta_uE[\boldsymbol{U}|\boldsymbol{A,BAV}]}{\partial \boldsymbol{A}}\nonumber\\
&= \beta_a + \beta_u\frac{\partial E[(\alpha_u + \zeta_a\boldsymbol{A} + \zeta_{bav} + \boldsymbol{\epsilon_1})|\boldsymbol{A,BAV}]}{\partial \boldsymbol{A}}\nonumber\\
&= \beta_a + \beta_u\zeta_a\nonumber\\
&= \beta_a + \beta_u\frac{\gamma_u\sigma_u^2}{\sigma_a^2 - \gamma_{bav}^2\sigma_{bav}^2}\label{Pearl Bias amp exp}
\end{align}
The approach of Pearl is limited in that it only works when the true underlying form of the conditional expectation, $E[\boldsymbol{U}| \boldsymbol{A, BAV}]$ is linear in both $\boldsymbol{A}$ and $\boldsymbol{BAV}$. As a result, the derivation is cumbersome and does not easily generalize to more complicated cases with more variables or different functional forms. Similarly we can find the expectation for the naive estimator $\naive$ from \eqref{seq naive} using this method, 

\begin{align}
   E[\naive] &= \frac{\partial E[\boldsymbol{Y}|\boldsymbol{A}]}{\partial \boldsymbol{A}}\\
   &= \frac{\partial E[\alpha_y + \boldsymbol{A}\beta_a + \boldsymbol{U}\beta_u +\boldsymbol{BAV}\beta_{bav} + \boldsymbol{\epsilon_1}|\boldsymbol{A}]}{\partial \boldsymbol{A}}\\
&= \beta_a + \frac{\partial \beta_uE[\boldsymbol{U}|\boldsymbol{A}]}{\partial \boldsymbol{A}} + \frac{\partial \beta_{bav}E[\boldsymbol{BAV}|\boldsymbol{A}]}{\partial \boldsymbol{A}}. \label{naive pearl exp expression}
\end{align}

We assume that the true underlying relationship between $\boldsymbol{U}$ and $\boldsymbol{A}$ is linear, while also assuming a linear relationship between $\boldsymbol{BAV}$ and $\boldsymbol{A}$:
\begin{align}
    \boldsymbol{U} &= \alpha_u + \boldsymbol{A}\tau_a + \boldsymbol{\epsilon_4}\label{linear U}\\
    \boldsymbol{BAV} &= \alpha_{bav} + \boldsymbol{A}\eta_a + \boldsymbol{\epsilon_5}\label{linear BAV}
\end{align}

Using these two equations (\ref{linear U}) and (\ref{linear BAV}), we arrive at the following expressions for the $COV(\boldsymbol{A,U})$ and $COV(\boldsymbol{A,BAV})$:

\begin{align}
    COV(\boldsymbol{A,U}) &= \tau_a\siga\\
    COV(\boldsymbol{A,BAV}) &= \eta_a\siga
\end{align}

Equation (\ref{cov au}) still holds from the above derivation. Following an analogous process, we can show that $COV(\boldsymbol{A,BAV}) = \gamma_{bav}\sigbav$ in terms of the original structural coefficients. Combining the four covariance expressions and solving for $\tau_a$ and $\eta_a$ in terms of the structural equations yields:

\begin{align}
    \tau_a &= \frac{\gamma_u\sigu}{\siga}\label{tau solve}\\
    \eta_a &= \frac{\gamma_{bav}\sigbav}{\siga}\label{eta solve}
\end{align}
Substituting (\ref{linear U}) and (\ref{linear BAV}) in (\ref{naive pearl exp expression}) and using (\ref{tau solve}) and (\ref{eta solve}) yields the following expectation for $\naive$:
\begin{align}
    E[\naive] = \beta_a + \beta_u\frac{\gamma_u\sigu}{\siga} + \beta_{bav}\frac{\gamma_{bav}\sigbav}{\siga}. 
\end{align}
Note in this derivation we needed to assume two linear relationships in order to derive the expectation, $E[\boldsymbol{U}|\boldsymbol{A}]$ and $E[\boldsymbol{BAV}|\boldsymbol{A}]$. These assumptions are not necessary when using probability limits to define limiting expressions in terms of the structural parameters.

\subsection{Probability Limit Calculations}\label{sbsec: plim calculations}

Now we will show the generality of probability limits for generating meaningful expressions of estimator behaviour and again we will use $\bav$ from equation \eqref{seq iv}, where $\boldsymbol{Z} = [\boldsymbol{\one},\boldsymbol{BAV}]$.\\

From the FWL theorem, $\bav = \frac{\boldsymbol{\amy}}{\boldsymbol{\ama}}$.

\begin{align}
\bav &= \frac{\boldsymbol{\amy}}{\boldsymbol{A^TM_zA}}\\
&= \frac{\boldsymbol{A^TM_z}(\ay + \boldsymbol{A}\ba + \boldsymbol{U}\bu + \boldsymbol{BAV}\bbav + \eone)}{\boldsymbol{\ama}}\\
&= \ba + \frac{\boldsymbol{A^TM_z}(\boldsymbol{U}\bu + \eone)}{\ama}\\
&= \ba + \bu\frac{\boldsymbol{\ama}}{\boldsymbol{\amu}} + \frac{\boldsymbol{\ama}}{\boldsymbol{A^TM_z\etwo}}\\
&= \ba + \bu\frac{\frac{1}{n}\boldsymbol{\amu}}{\frac{1}{n}\boldsymbol{\ama}} + \frac{\frac{1}{n}\boldsymbol{A^TM_z\etwo}}{\frac{1}{n}\boldsymbol{\ama}}\label{cond matrix form}
\end{align}
which follows by simply substituting in the true structural equations for $\boldsymbol{Y}$ and $\boldsymbol{A}$, \ref{Y truth} and \ref{A truth} respectively, and then applying the annihilating properties of $\boldsymbol{M_z}$ to set linear combinations of constants and $\boldsymbol{BAV}$ to 0. Notice that we have not used any information about structural equation for the treatment. Thus the numerical form in \eqref{cond matrix form} holds for any treatment structural equation, $\boldsymbol{A} = f(\boldsymbol{U,BAV,\epsilon_2})$. We use this result in section \ref{sbsec:non-linear}. Further, since this is written in general matrix notation, $\boldsymbol{U}$ and $\boldsymbol{BAV}$ can be trivially extended from vectors to be any finite dimension. This result is used in section \ref{sec:identify_amp}.\\

We can now solve for the probability limits of the three remaining expressions separately ($\frac{1}{n}\boldsymbol{\ama}$, $\frac{1}{n}\boldsymbol{\amu}$, and $\frac{1}{n}\boldsymbol{A^TM_z\etwo}$) and combine them due to the properties of probabilities limits, namely \eqref{plim add}, \eqref{plim mult}, and \eqref{plim div}. We begin with deriving $\frac{1}{n}\boldsymbol{A^TM_z\etwo}$),

\begin{align}
  \frac{1}{n}\boldsymbol{A^TM_z\etwo} &=\overset{p}\to E[\boldsymbol{A^TM_z\etwo}]\\
  &= E[\boldsymbol{A^TM_z}E[\boldsymbol{\etwo}|\boldsymbol{A,Z}]]\\
  &= E[\boldsymbol{A^TM_z}E[\boldsymbol{\etwo}]]\\
  &= 0, 
\end{align}
where the first line follows from the WLLN, line two from the Law of Iterated Expectations, and the third from the independence of $\boldsymbol{\etwo}$ from $\boldsymbol{A}$ and $\boldsymbol{BAV}$. Next we consider $\frac{1}{n}\boldsymbol{\ama}$,

\begin{align}
    \frac{1}{n}\boldsymbol{\ama} &= \frac{1}{n}(\alpha_a + \boldsymbol{U}\gu + \boldsymbol{BAV}\gbav + \boldsymbol{\etwo})^T\boldsymbol{M_z}(\alpha_a + \boldsymbol{U}\gu + \boldsymbol{BAV}\gbav + \etwo)\\
    &=  \frac{1}{n}(\boldsymbol{U}\gu + \etwo)^T\boldsymbol{M_z}(\boldsymbol{U}\gu + \etwo)\\
    &\overset{p}\to  \plim \frac{1}{n}(\boldsymbol{U}\gu + \etwo)^T\miota(\boldsymbol{U}\gu + \etwo)\label{m_z for m_1}\\ 
    &= \plim \frac{1}{n}(\boldsymbol{U}\gu)^T\miota(\boldsymbol{U}\gu) + \plim\frac{1}{n}2(\boldsymbol{U}\gu)^T\boldsymbol{\miota}\etwo + \plim\frac{1}{n}\etwo\boldsymbol{\miota}\etwo\\
    &= \gamma_u^2E[(\boldsymbol{U} - \bar{\boldsymbol{U}})^T(\boldsymbol{U}-\bar{\boldsymbol{U}})] + 2\gu E[\boldsymbol{U}E[\etwo|\boldsymbol{U}]] + E[\etwo^T\etwo]\\
    &= \gamma_u^2\sigu + 0 + \sigma_{\etwo}^2\\
    &= \siga - \gbav^2\sigbav, 
\end{align}

where line \ref{m_z for m_1} follows from the fact that $BAV$ is independent of both $U$ and $\etwo$. Since $M_z$ is a residual making vector, we can compare the residuals in the probability limit from the following two regressions:

\begin{align}
 (\boldsymbol{U}\gu + \etwo) = \alpha_{\boldsymbol{U}\gu + \etwo} + \boldsymbol{BAV}\eta_{bav} + \boldsymbol{\upsilon_1}\\
 (\boldsymbol{U}\gu + \etwo) = \alpha_{\boldsymbol{U}\gu + \etwo}+ \boldsymbol{\upsilon_2}
\end{align}

Due to independence, $\hat{\eta_{bav}} \overset{p}\to 0$ and thus the residuals from the two regressions will be equivalent asymptotically. Thus we can replace $\boldsymbol{M_z}$ with $\boldsymbol{\miota}$ in equation (\ref{m_z for m_1}), which as the centering projection matrix enjoys favorable properties as discussed in section \ref{sbsec: matrix and fwl}.\\

Finally we need to find the probability limit of $\frac{1}{n}\boldsymbol{\amu}$.

\begin{align}
  \frac{1}{n}\boldsymbol{\amu} &= \frac{1}{n}(\alpha_a + \boldsymbol{U}\gu + \boldsymbol{BAV}\gbav + \etwo)^T\boldsymbol{M_zU}\\
  &= \frac{1}{n}(\boldsymbol{U}\gu + \etwo)^T\boldsymbol{M_zU}\\
  &\overset{p}{\to} \frac{1}{n}(\boldsymbol{U}\gu + \etwo)^T\boldsymbol{\miota U}\\
  &= \frac{1}{n}(\gu(\boldsymbol{U^T\miota U}) + \gu \boldsymbol{U\miota} \etwo\\
  &\overset{p}{\to} \gu E[(\boldsymbol{U-\bar{U}})^2] + E[(\boldsymbol{U-\bar{U}})(\etwo - \bar{\etwo})]\\
  &= \gu\sigu. 
\end{align}

Putting this altogether this implies:

\begin{align}
    \bav \overset{p}{\to} \bu\frac{\gu\sigu}{\siga - \gbav^2\sigbav}
\end{align}

This is equivalent to the expectation in this case. The benefit is that it is more robust to functional form assumptions and by using properties (\ref{plim add})-(\ref{plim div}) and the FWL theorem we can find find asymptotic bias expressions by partitioning the estimator into a series of functions of residuals from simpler regressions. Further, we can always find the limiting expression for the numerator and the denominator separately. Expectations cannot be split up in such a manner and ratios of variables can be very difficult to find closed form expressions for the expectation without imposing restrictive assumptions.

\subsection{Additional Details Simulation} \label{sbsec: simulation invariant bav est}

Suppose we want to simulate a system of linear equations from equations \eqref{Y truth} and \eqref{A truth} based on the DAG in Figure \ref{fig:DAG1}. Now suppose we are interested in assessing the effect of modifying the edge $\boldsymbol{BAV} \to \boldsymbol{A}$ on the conditional estimator $\bav$. If we incorrectly run this simulation simply by changing the parameter $\gamma_{bav}$ to some (or some set of) $\gamma_{bav}^\prime$ and fail to fix the variance of the treatment $\boldsymbol{A}$ as discussed in section \ref{sc: new simulation}, we can show that the bias of estimator $\bav$ will remain unchanged. In section \ref{variance derivation} we show that the variance of $\boldsymbol{A}$ in the above simulation design is equal to:

\begin{align*}
    \sigma_a^2 = \gamma_u^2\sigma_u^2 + \gamma_{bav}^2\sigma_{bav}^2 + \sigma_{\epsilon_2}. 
\end{align*}
 Further we showed that the expectation and probability limit of the estimator is:
 \begin{align*}
     E[\bav] = \beta_a + \beta_u\frac{\gamma_u\sigma_u^2}{\sigma_a^2 - \gamma_{bav}^2\sigma_{bav}^2}. 
 \end{align*}
 
 Thus, if we change $\gamma_{bav}$ to $\gamma_{bav}^\prime$ holding all other parameters constant, it can be shown that the resulting expectation is unchanged. This is because the increased amplification is precisely cancelled out by increasing the variance of the treatment.
 
 \begin{align*}
     \sigma_a^\prime &= \gamma_u^2\sigma_u^2 + \gamma_{bav}^{\prime^2}\sigma_{bav}^2 + \sigma_{\epsilon_2}\\
     &= \sigma_a^2 - \gamma_{bav}^2\sigma_{bav}^2 + \gamma_{bav}^{\prime^2}\sigma_{bav}^2\\
     &= \sigma_a^2 + (\gamma_{bav}^{\prime^2} - \gamma_{bav}^2)\sigma_{bav}^2
 \end{align*}

This implies the expectation of the estimator $\bav^\prime$ has the following expression:
\begin{align*}
    E[\bav]^\prime &= \beta_a + \beta_u\frac{\gamma_u\sigma_u^2}{\sigma_a^{\prime^2} - \gamma_{bav}^{\prime^2}\sigma_{bav}^2}\\
    &= \beta_a + \beta_u\frac{\gamma_u\sigma_u^2}{(\sigma_a^2 + (\gamma_{bav}^{\prime^2} - \gamma_{bav}^2)\sigma_{bav}^2) - \gamma_{bav}^{\prime^2}\sigma_{bav}^2}\\
    &= \beta_a + \beta_u\frac{\gamma_u\sigma_u^2}{\sigma_a^2 - \gamma_{bav}^2\sigma_{bav}^2}\\
    &= E[\bav]
\end{align*}

Thus the expectation of the estimator remains unchanged for any change of parameter, $\gamma_{bav}$. As a consequence, the comparison of this estimator with the naive estimator will seem favorable as the absolute magnitude of the parameter $\gamma_{bav}^\prime$ increases since the difference in absolute bias is:
\begin{align*}
E[\naive] - E[\bav] = |\frac{\beta_u\gamma_u\sigma_u^2}{\sigma_a^2} + \frac{\beta_{bav}\gamma_{bav}\sigma_{bav}^2}{\sigma_a^2}| - |\beta_u\frac{\gamma_u\sigma_u^2}{\sigma_a^2 - \gamma_{bav}^2\sigma_{bav}^2}|
\end{align*}

The bias of $\naive$ is increasing in $\gamma_{bav}$ for sufficiently large $\gamma_{bav}$ and we showed above that the bias for $\bav$ is invariant to changes in $\gamma_{bav}$ if we do not fix the variance of the treatment $\boldsymbol{A}$. Thus eventually the bias of the naive estimate is strictly increasing in $\gamma_{bav}$ and will continue to appear worse and worse relative to the conditional estimator. However, as discussed in section \ref{sc: new simulation} this is a consequence of failing to conduct a proper causal simulation experiment comparing data sets plausibly generated from similar experiments and holding all other potentially confounding edges constant.  

\subsection{Real Data Simulation Details}\label{real sim details}

The goal of this section is to utilize the real randomized control trial data described in section \ref{sc:real data}, which comes from the DAG in Figure \ref{fig:DAG exp 1}, and simulate modified covariates ($\tilde{\boldsymbol{X}}$) and a modified outcome ($\tilde{Y}$) such that they come from the DAG \ref{fig:DAG exper 2} and the equations \ref{outcome mod rct}, \ref{latent mod rct}, and \ref{treatment mod rct}. This proceeds in two steps. First we need to simulate the latent variable $\boldsymbol{A^\star}$ and then conditionally simulate $\boldsymbol{BAV}$ and $\boldsymbol{U}$.\\

For simplicity, we set $\boldsymbol{U}$ and $\boldsymbol{BAV}$ to be standard normal variables. The latent variable, $\boldsymbol{A^\star}$, has variance of $1$ but its mean, $\alpha_a$ is determined to ensure that $P(\boldsymbol{A^\star} >0) = P(\boldsymbol{U}\gamma_u + \boldsymbol{BAV}\gamma_bav + \epsilon_2 > \alpha_a) = p_a$, which is determined in our data by matching $p_a$ to the observed quantity $\hat{p_a} = \frac{1}{n}\sum_{i=1}^n A_i \approx 0.51$.\\

Under the assumptions above, this implies that $\alpha_a = -\Phi^{-1}(1-p_a)$, where $\Phi(x)$ is the cumulative distribution function (CDF) of the standard normal distribution. As previously mentioned, the variance of the error term $\boldsymbol{\epsilon_2}$ is set precisely to ensure that variance of $\boldsymbol{A^\star}$ is 1, $\sigma_{\epsilon_2}^2 = 1 - \gamma_u^2 - \gamma_{bav}^2$.\\

 The first step is to use the observed $\boldsymbol{A}$ data to simulate the latent $\boldsymbol{A^\star}$. Consider the CDF of the latent $\boldsymbol{A^\star}$ conditional on $\boldsymbol{A}=1$. Let $P(\boldsymbol{A} =1) = p_a$, so that 

\begin{align*}
P(\boldsymbol{A^\star} \leq a^\star | \boldsymbol{A} =1) &=P(\boldsymbol{A^\star} \leq a^\star | \boldsymbol{A^\star} \geq 0)\\
&= \frac{P(\boldsymbol{A^\star} \leq a^\star, \boldsymbol{A^\star} \geq 0)}{P(\boldsymbol{A^\star} \geq 0)}\\
&= \frac{P(0 \leq \boldsymbol{A^\star} \leq a^\star)}{P(\boldsymbol{A^\star} \geq 0)}\\
&= \frac{P(\frac{-E[\boldsymbol{A^\star}]}{\sigma_{a^\star}^2} \leq \frac{\boldsymbol{A^\star}-E[\boldsymbol{A^\star}]}{\sigma_{a^\star}^2} \leq \frac{a^\star-E[\boldsymbol{A^\star}]}{\sigma_{a^\star}^2})}{P(\boldsymbol{A^\star} \geq 0)}\\
&= \frac{\Phi(\frac{a^\star-E[\boldsymbol{A^\star}]}{\sigma_{a^\star}^2}) - \Phi(\frac{-E[\boldsymbol{A^\star}]}{\sigma_{a^\star}^2})}{p_a}\\
&= \frac{\Phi(a^\star - \alpha_a) - \Phi(-\alpha_a)}{p_a}\\
&= \frac{\Phi(a^\star - \alpha_a) - \Phi(\Phi^{-1}(1-p_a))}{p_a}\\
&= \frac{\Phi(a^\star - \alpha_a) - (1-p_a)}{p_a}. 
\end{align*}
Since $P(\boldsymbol{A^\star} \leq a^\star | \boldsymbol{A} =1)$ is the CDF of a continuous random variable, it is distributed uniformly between 0 and 1. Let $\boldsymbol{X} \sim U(0,1)$ be a uniform random variable with support $[0,1]$. From the well-known Probability Inverse Transformation $P(\boldsymbol{A^\star} \leq a^\star | \boldsymbol{A} =1) \equiv X$, are equivalent in distribution.

\begin{align*}
&\implies \boldsymbol{X} \equiv \frac{\Phi(a^\star - \alpha_a) - (1-p_a)}{p_a}\\
&\implies p_a\boldsymbol{X} + (1-p_a) \equiv \Phi(a^\star - \alpha_a)\\
&\implies \Phi^{-1}( p_a\boldsymbol{X} + (1-p_a)) + \alpha_a \equiv a^\star
\end{align*}

Similarly, it can be shown that conditional on $A =0$:
\begin{align*}
a^\star \equiv \Phi^{-1}((1-p_a)\boldsymbol{X}) + \alpha_a. 
\end{align*}

Thus, in general:
\begin{align*}
    a^\star \equiv \Phi^{-1}(pa^{\boldsymbol{A}}(1-p_a)^{\boldsymbol{A}-1}\boldsymbol{X} +(1-p_a)^{\boldsymbol{A}}) + \alpha_a. 
\end{align*}

Therefore, by conditioning on $\boldsymbol{A}$ and simulating a uniform random variable,  we can take draws from the unobserved latent variable $\boldsymbol{A^\star}$. Once we have recovered the latent variable we can jointly simulate $\boldsymbol{U}$ and $\boldsymbol{BAV}$ conditional on $\boldsymbol{A^\star}$ and the observed covariates $\boldsymbol{X}$. The observed covariates are centered and  asympotically multivariate normal. Since $\boldsymbol{U}$, $\boldsymbol{BAV}$, and $\boldsymbol{A^\star}$ are univariate or multivariate normal variables and $\boldsymbol{X}$ are asymptotically normal, the conditional distribution will be asymptotically multivariate normal and proportional to the joint density. From standard multivariate normal theory:

\begin{align*}
\boldsymbol{U}=u, \boldsymbol{BAV} = bav &| \boldsymbol{A^\star} = a^\star, \boldsymbol{X} = x \sim N(\boldsymbol{\mu^\star},\boldsymbol{Sigma}^\star)
\end{align*}

As stated above the conditional distribution is proportional to the joint model. Thus we will define $\boldsymbol{\Sigma}$ and $\boldsymbol{\mu}$ for the joint density.

\begin{align*}
\boldsymbol{U}=u, \boldsymbol{BAV} = bav, \boldsymbol{A^\star} = a^\star, \boldsymbol{X} = x \sim N(\boldsymbol{\mu},\boldsymbol{\Sigma})
\end{align*}

\begin{align*}
\boldsymbol{\Sigma} &= \begin{bmatrix}
\boldsymbol{\Sigma_{(U,BAV),(U,BAV)}} & \boldsymbol{\Sigma_{(U,BAV),(A^\star,X)}}\\
\boldsymbol{\Sigma_{(A^\star,X),(U,BAV)}} & \boldsymbol{\Sigma_{(A^\star,X),(A^\star,X)}}
\end{bmatrix}
\end{align*}

\begin{align*}
\boldsymbol{\Sigma_{(U,BAV),(U,BAV)}} &= \begin{bmatrix}
\sigma_u^2 & 0 & 0 & 0\\
0 & \sigma_{bav_1}^2 &0 &0\\
0& 0 & \sigma_{bav_2}^2 &0\\
0 & 0 & 0 & \sigma_{bav_3}^2
\end{bmatrix}
\end{align*}

\begin{align*}
\boldsymbol{\Sigma_{(U,BAV),(A^\star,X)}} &= \begin{bmatrix}
\gamma_u\sigma_u^2 & 0 & 0 & 0\\
\gamma_{\Tilde{x_1}}\sigma_{bav_1}^2 & 0 &0 &0\\
\gamma_{\Tilde{x_2}}\sigma_{bav_2}^2& 0 & 0 &0\\
\gamma_{\Tilde{x_3}}\sigma_{bav_3}^2 & 0 & 0 & 0
\end{bmatrix}
\end{align*}

\begin{align*}
\boldsymbol{\Sigma_{(A^\star,X),(U,BAV)}} &= \boldsymbol{\Sigma_{(U,BAV),(A^\star,X)}^T}
\end{align*}

\begin{align*}
   \boldsymbol{\Sigma_{(A^\star,X),(A^\star,X)}} &=  \begin{bmatrix}
\sigma_{a^\star}^2 & 0 & 0 & 0\\
0 & \sigma_{x_1}^2 &0 &0\\
0& 0 & \sigma_{x_2}^2 &0\\
0 & 0 & 0 & \sigma_{x_3}^2
\end{bmatrix}
\end{align*}

\begin{align*}
    \boldsymbol{\mu} &= \begin{bmatrix}
    \mu_u\\
    \mu_{bav_1}\\
    \mu_{bav_2}\\
    \mu_{bav_3}\\
    \mu_{a^\star}\\
    \mu_{x_1}\\
    \mu_{x_2}\\
    \mu_{x_3}
    \end{bmatrix}
    = \begin{bmatrix}
    0\\
    0\\
    0\\
    0\\
    \alpha_a\\
    0\\
    0\\
    0
    \end{bmatrix}. 
\end{align*}

Using standard multivariate normal theory and the matrices defined above we can define the mean, $\boldsymbol{\mu^\star}$ and variance, $\boldsymbol{\Sigma^\star}$ of the conditional model: 

\begin{align*}
    \boldsymbol{\Sigma^\star} = \boldsymbol{\Sigma_{(U,BAV),(U,BAV)}} - \boldsymbol{\Sigma_{(U,BAV),(A^\star,X)}}\boldsymbol{\Sigma_{(A^\star,X),(A^\star,X)}^{-1}}\boldsymbol{\Sigma_{(A^\star,X),(U,BAV)}}; \mbox{ and }
\end{align*}

\begin{align*}
    \boldsymbol{\mu^\star} = \boldsymbol{\Sigma_{(U,BAV),(A^\star,X)}\Sigma_{(A^\star,X),(A^\star,X)}^{-1}}([\boldsymbol{A^\star,X}] - \begin{bmatrix} \mu_{a^\star}\\
    \boldsymbol{\mu_x}
    \end{bmatrix})^T. 
\end{align*}

Using the conditional distribution, we can thus take draws of $\boldsymbol{U,BAV}$ conditional on the particular values of $\boldsymbol{A^\star}$ and $\boldsymbol{X}$. Using the simulated $\boldsymbol{BAV}$ we add it to the covariates $\boldsymbol{X}$ to form the modified covariates, $\boldsymbol{\tilde{X}} = \frac{\boldsymbol{X}}{\sigma^\prime} + \boldsymbol{BAV}$, where $\sigma^\prime$ is a scaling factor chosen simulataneously with $\sigma_{bav}$ such that the variance of $\boldsymbol{\tilde{X}}$ is precisely equal to $\boldsymbol{\sigma_{x}} =1$. This step is important if we would like to compare simulations with the modified and the unmodified covariates. In the particular simulations conducted in section \ref{sc:real data}, the scaling was chosen such that $Var(\frac{\boldsymbol{X_i}}{\sigma^\prime}) = 0.01, i = 1,2,3$ and thus $\sigma_{bav_i}^2 = 0.99, i = 1,2,3$.\\

Now that the modified covariates have been constructed, the modified outcome can be constructed. The original RCT data coming from Figure \ref{fig:DAG exp 1} is assumed to come from the linear model:
\begin{align*}
    \boldsymbol{Y} = \alpha_y + \boldsymbol{A}\beta_a + \boldsymbol{X}\beta_x + \epsilon_1, 
\end{align*}

\noindent where $\beta_a$ and $\beta_x$ are estimated unbiasedly in section \ref{sc:real data}. Next, we add the unmeasured confounding, $\boldsymbol{U}\beta_u$ directly (where $\beta_u$ is chosen) and then add $\boldsymbol{BAV}\beta_x$ + $\boldsymbol{\tilde{X}}\beta_{adj}$, where $\beta_{adj} = \beta_{\tilde{x}} - \beta_x$ where $\beta_{\tilde{x}}$ is chosen to set the desired covariance between the modified covariates and the modified outcome, 
\begin{align*}
   \boldsymbol{\tilde{Y}} &= \boldsymbol{Y} + \boldsymbol{U}\beta_u + \boldsymbol{BAV}\beta_x + \boldsymbol{\tilde{X}}(\beta_{\tilde{x}} - \beta_x)\\
   &= (\alpha_y + \boldsymbol{A}\beta_a + \boldsymbol{X}\beta_x + \boldsymbol{\epsilon_1}) + \boldsymbol{U}\beta_u + \boldsymbol{BAV}\beta_{x} + \boldsymbol{\tilde{X}}(\beta_{\tilde{x}} - \beta_x)\\
   &= \alpha_y + \boldsymbol{A}\beta_a + \boldsymbol{U}\beta_u + (\boldsymbol{X} + \boldsymbol{BAV})\beta_x + \boldsymbol{\tilde{X}}(\beta_{\tilde{x}} - \beta_x)+ \boldsymbol{\epsilon_1}\\
   &= \alpha_y + \boldsymbol{A}\beta_a + \boldsymbol{U}\beta_u + (\boldsymbol{\tilde{X}})\beta_x + \boldsymbol{\tilde{X}}(\beta_{\tilde{x}} - \beta_x)+ \boldsymbol{\epsilon_1}\\
   &= \alpha_y + \boldsymbol{A}\beta_a + \boldsymbol{U}\beta_u + \boldsymbol{\tilde{X}}\beta_{\tilde{x}}+ \boldsymbol{\epsilon_1}, 
\end{align*}

This is precisely the outcome equation in equation \ref{outcome mod rct} in Section \ref{sbsec: real data sim main}. Thus following this method we can use the real data to create a data simulation using the original treatment data and matching many of the characteristics of the real data, but precisely control the causal structure and correlations between the variables. As with the other simulations, there will still be restrictions on the parameters and correlations that we set such as positive definiteness of all the variance matrices in the above simulation.

\subsection{Real Data Simulation Comparison of Estimators}\label{sbsec: real sim estimator}

Consider a causal simulation experiment coming from a DAG and system of equations identical to the one considered in section \ref{sbsec: real data sim main} as described by Figure \ref{fig:DAG exper 2} and the system of equations \eqref{outcome mod rct}, \eqref{latent mod rct}, and \eqref{treatment mod rct}. The experiment uses the real data described in section \ref{sc:real data} and the procedure detailed in section \ref{real sim details}. The simulation experiment involves intervening on the edge $\boldsymbol{\tilde{X_1}}\to \boldsymbol{A}$, that is increasing the covariance between $\boldsymbol{\tilde{X_1}}$ and $\boldsymbol{A}$. As in section \ref{sc: new simulation} we will explore the consequences of failing to properly hold all non-intervention edges of the DAG.\\

\begin{table}[H]
\centering
\begin{tabular}{@{}cccccl@{}}
\toprule
Simulation Parameters                         & $\gamma_{\tilde{X}}$                  & $\gamma_u$                & $\beta_u$                 & $\boldsymbol{\beta_{\tilde{x}}}$                     & \multicolumn{1}{c}{$\beta_{a}$} \\ \midrule
\multicolumn{1}{c|}{Control}      & \multicolumn{1}{c|}{0.20, 0.38, 0.33} & \multicolumn{1}{c|}{0.63} & \multicolumn{1}{c|}{0.15} & \multicolumn{1}{c|}{0.10, -0.15, -0.10} & 0.1377                          \\
\multicolumn{1}{c|}{Intervention} & \multicolumn{1}{c|}{0.55, 0.38, 0.33} & \multicolumn{1}{c|}{0.63} & \multicolumn{1}{l|}{0.15} & \multicolumn{1}{c|}{0.10, -0.15, -0.10} & 0.1377                          \\ \bottomrule
\end{tabular}
\end{table}

Above, the parameters for the two simulation treatments are described. The only difference between the two is that in the control, $\gamma_{\tilde{x_1}} = 0.2$ and in the intervention $\gamma_{\tilde{x_1}} = 0.55$. Below we visualize the naive, adjusted, and unbiased estimators for the control treatment.

\begin{figure}[H]
\centering
\includegraphics[scale=.4]{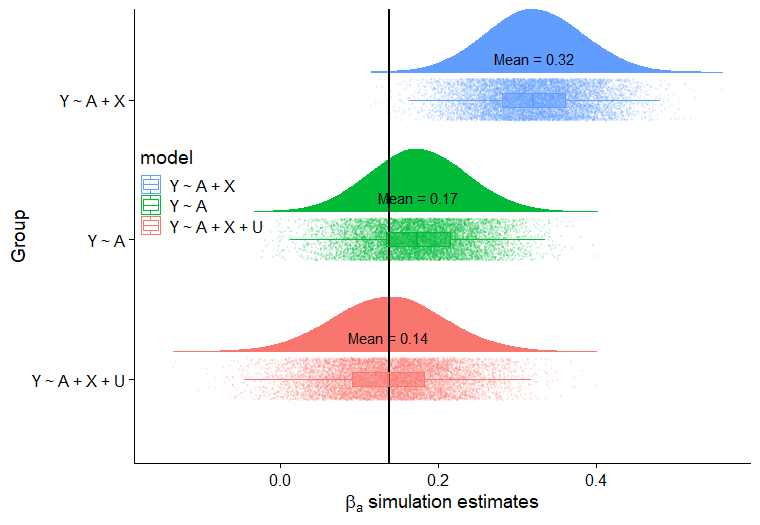}
\caption{Control treatment estimators. Black line represents the true underlying parameter $\beta_a = .1377$. $N = 10000$ simulation replications.}
\end{figure}

In the control treatment we see that the unbiased estimator behaves as expected, centered on the true underlying parameter. The naive estimator, $\naive$ is only slightly biased, since some unmeasured biases due to the vector $\boldsymbol{\tilde{X}}$ and $\boldsymbol{U}$ happen to be of opposing signs and partially cancel each other out. If this is not the case, of course the naive estimator may be significantly more biased. The adjusted estimator behaves poorly with an average absolute bias of $0.18$. Although the parameters in the latent space are relatively large, $\boldsymbol{\gamma_{\tilde{x}}} = [0.2, 0.38, 0.33]$, the covariances in the observed space with respect to the treatment are relatively small, $COV(\boldsymbol{A,\tilde{X}}) = [0.08,0.15,0.13]$, and yet the amplifying effect is quite large. In fact, the amplifying variables jointly explain only $18\%$ of the variance of the treatment, but since the variance of the treatment was already quite small, $\sigma_a^2 \approx 0.25$, the amplifying variables had a more than proportional effect.\\

The bias attributed to the path $\boldsymbol{A}\leftarrow \boldsymbol{U} \to \boldsymbol{Y}$ for the naive estimator is $\frac{\beta_uCOV(\boldsymbol{A,U})}{\sigma_a^2} = \frac{0.15\times 0.25}{0.25} = \frac{0.0375}{.25} = 0.0375\times 4 = 0.15$, whereas for the amplified estimator it is $\frac{0.15\times 0.25}{0.25 - (0.08^2 + 0.15^2 + 0.13^2)} = \frac{0.0375}{.183} = 0.0375\times 4.88 = 0.183$. Since $\boldsymbol{A^TMzA} \leq \boldsymbol{A^TM_{\iota}A}, \forall \boldsymbol{Z}: \boldsymbol{\iota} \subseteq \boldsymbol{Z}$, we can rewrite the bias due to bias amplification as $\frac{|\beta_u\times COV(\boldsymbol{A,U})|}{(1-c)\times \sigma_a^2}, c\in [0,1]$, where $c$ is the proportion of treatment variance explained by the bias amplifiers jointly, 

\begin{align*}
\frac{\partial^2( \frac{|\beta_u\times COV(\boldsymbol{A,U})|}{(1-c)\times \sigma_a^2})}{\partial \sigma_a^2 \partial c} &= \begin{cases}
& \frac{-\beta_u\times COV(\boldsymbol{A,U})}{(1-c)^2(\sigma_a^2)^2}, \beta_u\times COV(\boldsymbol{A,U}) > 0\\
& \frac{\beta_u\times COV(\boldsymbol{A,U})}{(1-c)^2(\sigma_a^2)^2}, \beta_u\times COV(\boldsymbol{A,U}) <0. 
\end{cases}
\end{align*}

The derivative above shows us that as the variance, $\sigma_a^2$, gets smaller, the marginal impact on absolute bias from an increase in the proportion of the variance explained by the amplifiers increases.\\

Now consider the intervention of increasing the proportion of variance explained by one of the potential amplifiers, $\boldsymbol{\tilde{X_1}}$, by increasing $\gamma_{\tilde{x_1}}$ to $0.55$. Again, we will consider the case of keeping all of the variances constant to the case where we simply change the parameter and allow the variances to float.

\begin{figure}[H]
    \centering
    \includegraphics[scale=0.6]{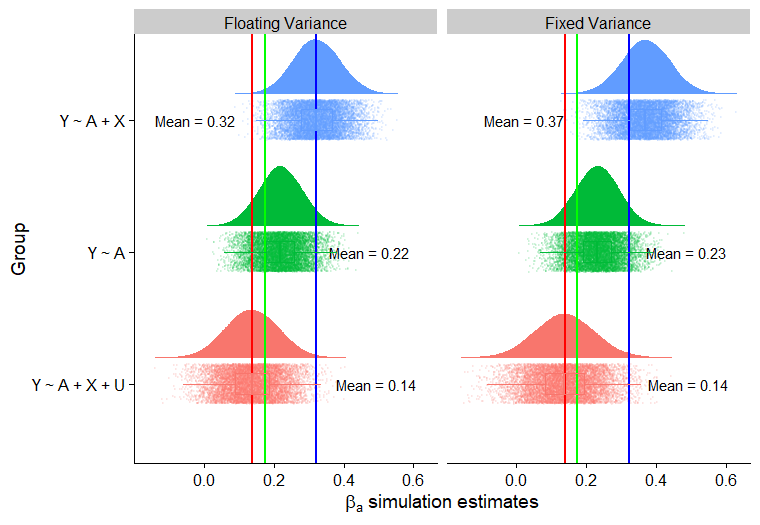}
    \caption{The vertical lines represent the means from the control experiment: red representing the mean of unbiased control estimator, green the mean of the naive estimator, and blue the mean of the amplified estimator}
    \label{fig:raincloud facet}
\end{figure}

Notice that in the left panel that although we have intentionally increased the amplification, the amplifying estimator has seemingly not changed. However, when we fix the variance, the bias amplification increases as we expected (the mean absolute bias increased from $0.18$ to $0.23$). The reason for this effect is that as the variance of the latent treatment $\boldsymbol{A^\star}$ increases, the covariances of the variables of the unmeasured confounding and the treatment as well as the potential amplifiers, $COV(\boldsymbol{A,U})$ and $COV(\boldsymbol{A,\tilde{X}})$, decrease. It can be shown that when $\boldsymbol{A^\star}, \boldsymbol{\tilde{X}},$ and $\boldsymbol{U}$ are normal or multivariate normal that:

\begin{align}
    COV(\boldsymbol{A,U}) &= \frac{1}{p_a}E[\boldsymbol{U}] + \frac{\gamma_u\sigma_u^2}{\sqrt{2\pi\sigma_{a^\star}^2}}\exp(\frac{-\alpha_a^2}{2\sigma_{a^\star}^2})\label{cov(a,u)bin}\\
        COV(\boldsymbol{A,\tilde{X}}) &= \frac{1}{p_a}E[\boldsymbol{\tilde{X}}] + \frac{\boldsymbol{\gamma_{\tilde{x}}}\boldsymbol{\sigma_{\tilde{x}}^2}}{\sqrt{2\pi\sigma_{a^\star}^2}}\exp(\frac{-\alpha_a^2}{2\sigma_{a^\star}^2})\label{cov(u,x)}. 
\end{align}

We can see in equation \eqref{cov(a,u)bin} that if we allow the variance of $\boldsymbol{A^\star}$ to increase as $\gamma_{\tilde{x_1}}$ increases that $COV(\boldsymbol{A,U})$ decreases. In the case of this simulation, the covariance decreased from $0.25$ in the control treatment to $0.22$, since $\sigma_{a^\star}^2$ increased from $1$ to $1 + (0.55^2 - 0.2^2) = 1.26$. Thus we have decreased the strength of the edge $\boldsymbol{U}\to \boldsymbol{A}$ incidentally. Further by considering equation \eqref{cov(u,x)} we can see that if we increase $\gamma_{\tilde{x_1}}$ we do not necessarily increase the amount of variance explained by $\boldsymbol{\tilde{X_1}}$ since there are two opposing effects. First, consider the increase directly through $\gamma_{\tilde{x_1}}$ and the decrease through increasing $\sigma_{a^\star}^2$. In the particular example, although our intended goal was to observe the effect of increasing the weight of the edge $\boldsymbol{\tilde{X_1}} \to \boldsymbol{A}$ we have in fact inadvertently decreased the covariance from $0.2$ to $0.196$.\\

Again we can see that when we fail to hold the variances constant, we are no longer comparing a controlled intervention on the weight of a particular set of nodes, but have modified the edges into and out of the intervened upon edge. This example shows that this is true in cases beyond fully linear systems of equations explored in section \ref{sc: new simulation}. Examining the simulation results we can see that this might lead to inappropriate conclusions about the effects of our interventions and the relative merits of particular estimators in contexts of interest to us.

\subsection{Simulation for Figure \ref{biasviz}}
\label{sbsec:simfig}
The Figure was simulated from the general structural equations (\ref{Y truth}) and (\ref{A truth}) with the particular values below.
  \begin{align*}
\boldsymbol{Y} &= 2 + .2\times \boldsymbol{A} + .5\times \boldsymbol{U} + .05\times \boldsymbol{BAV} + \boldsymbol{\upsilon_1}\\
\boldsymbol{A} &= 1 + .3\times \boldsymbol{U} + .75\times \boldsymbol{BAV} + \boldsymbol{\upsilon_2}\\
\boldsymbol{BAV} &\sim N(0,1), \quad \boldsymbol{U} \sim N(0,1)\\
\boldsymbol{\upsilon_1} &\sim N(0, \sigma_{\upsilon_1})\\
\sigma_{\upsilon_1} &= (\sigma_y^2 - (\beta_a^2\sigma_a^2 + \beta_u^2\sigma_u^2 + \beta_{bav}^2\sigma_{bav}^2\\
&+ 2\beta_a\beta_u\gamma_u\sigma_u^2 + 2\beta_a\beta_{bav}\gamma_{bav}\sigma_{bav}^2\sigma_{\epsilon_1}^2))^{\frac{1}{2}}\\
&= 0.906\\
\boldsymbol{\upsilon_2} &\sim N(0,\sigma_{\upsilon_2})\\
\sigma_{\upsilon_2} &= (\sigma_a^2 -(\gamma_u^2\sigma_u^2 + \gamma_{bav}^2\sigma_{bav}^2 +\sigma_{\epsilon_2}^2))^{\frac{1}{2}}\\
&= 0.809
\end{align*}

$\boldsymbol{\upsilon_1}$ and $\boldsymbol{\upsilon_2}$ had variances such that A and Y both have unit variance.

\subsection{Variance Derivations}\label{variance derivation}

\paragraph{Treatment Variance for equation (\ref{A truth})}
\begin{align}
\boldsymbol{A} &= \alpha_a + \boldsymbol{U}\gamma_u + \boldsymbol{BAV}\gamma_{bav} + \boldsymbol{\epsilon_2} \nonumber\\
\implies Var(\boldsymbol{A}) &= \gamma_u^2\sigma_u^2 + \gamma_{bav}^2\sigma_{bav}^2 + 2\gamma_u\gamma_{bav}COV(\boldsymbol{U,BAV}) + Var(\boldsymbol{\epsilon_2})\nonumber\\
\sigma_a^2 &= \gamma_u^2\sigma_u^2 + \gamma_{bav}^2\sigma_{bav}^2 +\sigma_{\epsilon_2}^2
\end{align}

\paragraph{Outcome Variance for equation (\ref{Y truth})}
\begin{align}
    \boldsymbol{Y} &= \alpha_y + \boldsymbol{A}\ba + \boldsymbol{U}\bu + \boldsymbol{BAV}\bbav + \boldsymbol{\eone}\\
    \implies Var(\boldsymbol{Y}) &= \beta_a^2Var(\boldsymbol{A}) + \beta_u^2Var(\boldsymbol{U}) + \beta_{bav}Var(\boldsymbol{BAV}) + \sigma_{\etwo}^2+ \\
    & 2\ba\bu Cov(\boldsymbol{A,U}) + 2\ba\bbav Cov(\boldsymbol{A,BAV}) + 2\bu\bbav Cov(\boldsymbol{U,BAV})\\
    &= \beta_a^2\siga + \beta_u^2\sigu + \beta_{bav}^2\sigbav + \sigma_{\etwo}^2 + 2\beta_{a}\bu\gamma_u\sigu + 2\ba\bbav\gbav\sigbav
\end{align}

\end{document}